\documentclass[iop]{emulateapj}
\usepackage{amsmath}

\slugcomment{Accepted on June 3, 2011 for Publication in Astrophysical Journal Supplements}

\shorttitle{Bulge+Disk Decompositions of SDSS Galaxies}
\shortauthors{Simard et al.}

\def\littleprime{\ifmmode{\scriptscriptstyle \prime }   
\else{\hbox{$\scriptscriptstyle \prime$ }}\fi}

\def\arcsec{\raise .9ex \hbox{\littleprime\hskip-3pt\littleprime}}
\def\arcmin{\raise .9ex \hbox{\littleprime}}
\def\arcsecpoint{\hbox to 1pt{}\rlap{\arcsec}.\hbox to 2pt{}}
\def\arcminpoint{\hbox to 1pt{}\rlap{\arcmin}.\hbox to 2pt{}}

\usepackage{graphics}

\begin{document}

\title{A Catalog of Bulge+Disk Decompositions and Updated Photometry for 1.12 Million Galaxies in the Sloan Digital Sky Survey}

\author{Luc Simard\altaffilmark{1,2}, J. Trevor Mendel\altaffilmark{2}, David R. Patton\altaffilmark{3}, Sara L. Ellison\altaffilmark{2}, Alan W. McConnachie\altaffilmark{1}}

\altaffiltext{1}{National Research Council of Canada, Herzberg Institute of Astrophysics, 5071 West Saanich Road, Victoria, British Columbia, Canada}
\altaffiltext{2}{Department of Physics and Astronomy, University of Victoria, Victoria, British Columbina, V8P 1A1, Canada}
\altaffiltext{3}{Department of Physics and Astronomy, Trent University, 1600 West Bank Drive, Peterborough, Ontario, K9J 7B8, Canada}

\begin{abstract}
We perform two-dimensional, Point-Spread-Function-convolved, bulge+disk decompositions in the
$g$ and $r$ bandpasses on a sample of 1,123,718  galaxies from the Legacy area of the Sloan Digital Sky Survey Data Release Seven. Four different decomposition procedures are investigated which make improvements to sky background determinations and object deblending over the standard SDSS procedures that lead to more robust structural parameters and integrated galaxy magnitudes and colors, especially in crowded environments. We use a set of science-based quality assurance metrics namely the disk luminosity-size relation, the galaxy color-magnitude diagram and the galaxy  central (fiber) colors  to show the robustness of our  structural parameters. The best procedure utilizes simultaneous, two-bandpass decompositions. Bulge and disk photometric errors remain below 0.1 mag down to bulge and disk magnitudes of  $g \simeq 19$ and $r \simeq 18.5$. We also use and compare three different galaxy fitting models: a pure S\'ersic model, a $n_b=4$ bulge + disk model and a S\'ersic (free $n_b$) bulge + disk model. The most appropriate model for a given galaxy is determined by the $F$-test probability. All three catalogs of measured structural parameters, rest-frame magnitudes and colors are publicly released here. These catalogs should provide an extensive comparison set for a wide range of observational and theoretical studies of galaxies.
\end{abstract}


\section{Introduction}\label{intro}
The properties of disks and spheroids in the local Universe are the direct outcomes of the hierarchical assembly of galaxies over cosmological time. It is well known that the structural parameters of these important galaxy sub-components obey scaling relations between size, luminosity and internal velocity \citep[e.g.,][]{courteau07}. The amount of stellar mass found in disks and bulges places strong constraints on the galaxy merger tree from $\Lambda$CDM N-body simulations \citep{hopkins10} and the secular formation of bulges \citep[e.g.,][]{kormendy79,athanassoula03}. Some properties such as disk size can be traced to properties of host dark matter haloes such as specific angular momentum \citep{fall80,mmw98}, and more information is needed for a truly large sample of galaxies to improve the recipes used in semi-analytical approaches to the treatment of the so-called ``sub-grid" physics \citep{governato04,robertson04,dutton09}.

Thanks to the availability of large datasets and powerful computing clusters, it is now possible to analyze large samples of galaxies in a fully quantitative way. Quantitative measurements of galaxy properties have made significant progress over the last few years \citep{peng02,simard02,desouza04,lotz04,conselice06,pignatelli06} and now provide an important framework for comparisons between observation and theory. A number of previous works have looked at the morphologies of SDSS galaxies using a wide range of subsamples and techniques \citep{shen03,blanton05a,kelly05,benson07,fukugita07,lintott08,labarbera10,wijesinghe10}. This paper presents the largest catalog of bulge+disk structural parameters measured to date of galaxies in the Sloan Digital Sky Survey Data Release Seven \citep[SDSS DR7,][]{abazajian09}. This catalog has already been used to study visual versus quantitative morphologies \citep{cheng10}, galaxy pairs \citep{ellison08,ellison10, patton11}, disk scaling relations at low and high redshift \citep{dutton11a,dutton11b}, the disk size function \citep[][and Simard 2011, in preparation]{kanwar08}, the evolution of cluster galaxies with redshift \citep{simard09}, compact groups of galaxies \citep{mendel11}, and the disk and bulge stellar mass functions \citep{thanjavur11}. The paper is organized as follows. The data and the bulge+disk decompositions are described in Sections~\ref{data} and~\ref{analysis}.  The catalog, science-based quality assessment metrics and a comparison between different fitting models are presented in Section~\ref{results}. The cosmology adopted throughout this paper is ($H_0, \Omega_{m}, \Omega_{\Lambda}$) = (70 km s$^{-1}$ Mpc$^{-1}$, 0.3, 0.7).

\section{Data}\label{data}

The data come from the Legacy area of the SDSS. We selected a photometric sample from the {\tt PhotoPrimary} table. The two main galaxy selection criteria were Petrosian magnitude (corrected for Galactic extinction) and morphological classification. We first selected objects with $ 14 \leq m_{petro,r,corr} \leq 18$, where $m_{petro,r,corr}$ is the $r$-band Petrosian magnitude corrected for Galactic extinction according to the extinction values given  in the SDSS database. We also selected extended objects with morphological type {\tt Type=3}. In addition to these two criteria, we also required that the sum of the flags DEBLENDED\_AS\_PSF and SATURATED be zero to eliminate objects that were found to be unresolved children of their parents as well as saturated objects. The full query reads: 

{\tt select objid from dr7.PhotoPrimary where flags \& (dbo.fPhotoFlags('SATURATED') + dbo.fPhotoFlags('DEBLENDED\_AS\_PSF'))= 0 and (petroMag\_r-extinction\_r) between 14.0 and 18.0 and Type=3}. 

\noindent This query returned 2,195,875 objects from the SEGUE and Legacy areas. In order to select only objects from the Legacy area, we used the DR7 sky coverage table {\tt allrunsdr7db.par}\footnote{http://www.sdss.org/dr7/coverage/allrunsdr7db.par} to obtain the list of run/rerun combinations in this area, and we then filtered the output of the query using the list. The total number of objects in our Legacy-area photometric sample is 1,123,718. 

\begin{figure}
\includegraphics[angle=270,width=8.7cm]{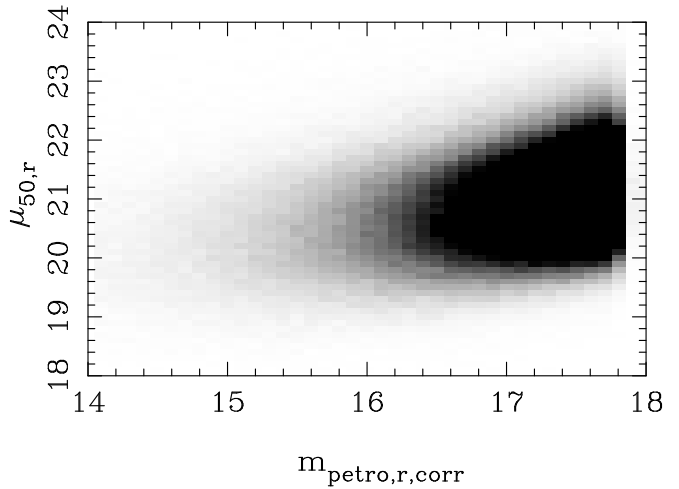}
\caption{Apparent $r$-band Petrosian magnitude and surface brightness distribution of galaxies in our photometric sample. Corrected for Galactic extinction using SDSS values.}
\label{petromag_mu50}
\end{figure}

Given the importance of the links between the morphological and spectroscopic properties of galaxies, we also defined a spectroscopic galaxy sample by selecting objects from our photometric sample with $m_{petro,r,corr} \leq 17.77$ and the spectrum of a galaxy as defined by the keyword {\tt SpecClass} in the {\tt SpecPhoto} database table. Specifically, we selected objects with {\tt SpecPhoto.SpecClass=2}.  These spectroscopic selection criteria yielded  674,701 galaxies. The faint limit is the completeness limit of the SDSS Main Galaxy spectroscopic sample \citep{strauss02}. As discussed in \citet{strauss02}, the nominal surface brightness limit of the SDSS spectroscopic sample is $r$-band $\mu_{50,r}$ = 24.5 mag arcsec$^{-2}$.  However, objects in the range 23.0 mag arcsec$^{-2}$ $\leq \mu_{50,r} \leq$ 24.5 mag arcsec$^{-2}$ were targeted only when the local and global sky values were within 0.05 mag arcsec$^{-2}$ of one another. We therefore set the faint surface brightness limit of our spectroscopic sample to $\mu_{50,r}$ = 23.0 mag arcsec$^{-2}$ to retain a complete sample. This limit further cuts the number of galaxies down to 671,425. The number of objects excluded by this cut is small as expected from Figure~\ref{petromag_mu50}. Following \citet{shen03}, we then excluded a small number of galaxies with redshift $z < 0.005$ whose distances may be severely contaminated by their peculiar velocities. The final total number of objects satisfying our spectroscopic selection criteria was 670,131. 

\begin{figure}
\includegraphics[angle=270,width=8.7cm]{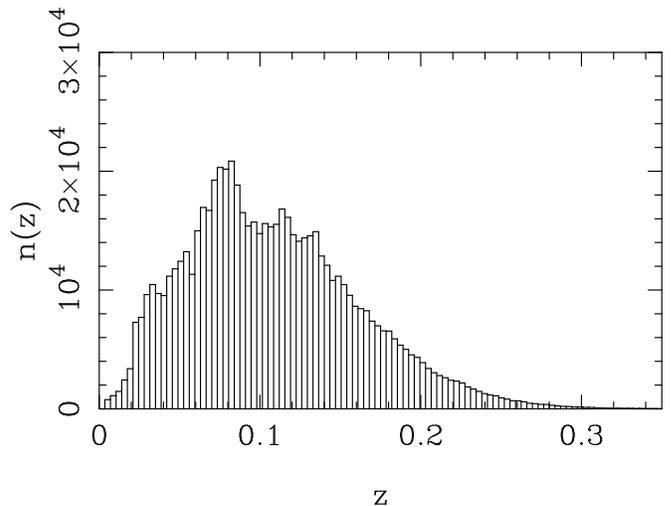}
\caption{Redshift distribution of galaxies in our spectroscopic sample ($14 \leq m_{petro,r,corr} \leq 17.77$, $\mu_{50,r} \leq 23$ mag arcsec$^{-2}$, {\tt SpecPhoto.SpecClass=2}).}
\label{zhist}
\end{figure}

$g$- and $r$-band images of all galaxies in our photometric sample were analyzed as described in Section~\ref{analysis}. The redshift distribution for the spectroscopic sample is shown in Figure~\ref{zhist}. The bulk of the sample is at $z \sim $ 0.1. The peaks in this distribution are due to real large-scale structures within the SDSS survey volume.

\section{Analysis}\label{analysis}

\subsection{Preprocessing of SDSS Images}\label{preproc}
Information for each galaxy to be analyzed was fetched directly from
the SDSS catalog server using a remote data mining pipeline based on the Python programming language and the {\tt wget} protocol. The fields {\tt objID}, {\tt ra}, {\tt dec},
{\tt run}, {\tt rerun}, {\tt camcol}, {\tt field}, {\tt obj}, {\tt
petroMag\_r}, {\tt z}, and {\tt extinction\_r} were extracted from the
{\tt PhotoPrimary} table. The atlas (prefix ``fpAtlas''), corrected
(prefix ``fpC'') and point-spread-function (prefix ``psField'') images
were then retrieved directly from the SDSS Data Archive Server (DAS). The locations
of these images for a given galaxy were given by {\tt run}, {\tt
rerun} and {\tt camcol}. Galaxy positions ({\tt rowc} and {\tt colc})
on the corrected images were also extracted from {\tt PhotoPrimary}. The SDSS corrected images are survey images that have been bias subtracted, flat-fielded and purged of bright stars.

Our bulge+disk decomposition procedure requires input thumbnail images for each object. These
thumbnails were prepared from SDSS images using the following
pre-processing steps: (1) extraction of the atlas and
point-spread-function images, (2) creation of the science thumbnail
images, and (3) creation of the mask thumbnail image. We used the SDSS
software utilities {\tt
readAtlasImages}\footnote{http://www.sdss.org/dr7/products/images/read\_atlas.html} and {\tt
read\_PSF}\footnote{http://www.sdss.org/dr7/products/images/read\_psf.html}
for the atlas and PSF images respectively. {\tt readAtlasImages} requires a
bandpass and object ID as inputs while {\tt read\_PSF} can output a
SDSS-provided PSF at any object position (given by {\tt rowc}, {\tt colc}) in a
given bandpass.

\subsection{Defining Objects on SDSS images}\label{defining-objects}
\begin{figure*}
\begin{center}
\includegraphics[angle=270,width=12cm]{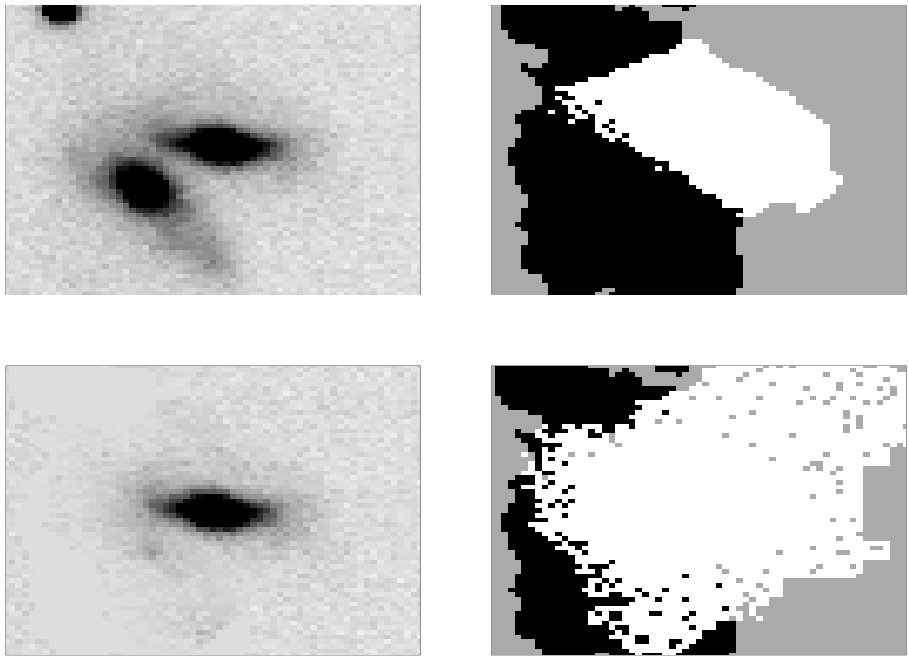}
\caption{The difference between SDSS and SExtractor deblending for object SDSS ObjID = 587729781199864370. This is the central galaxy in the image, and it has a close companion to the lower left. {\it Top left}: Object image from the SDSS corrected frame. {\it Top right}: Segmentation pixel mask produced by SExtractor as described in Section~\ref{sex-deblend}. The white region is the area defined as the object by the SExtractor deblending. {\it Bottom left}: Object image given by the SDSS deblender. {\it Bottom right}: Segmentation pixel mask constructed from the SDSS deblending as described in Section~\ref{sdss-deblend}. The white region is the area defined as the object by the SDSS atlas image. Note how it now extends over the area where the close companion used to be.}
\label{deblending-comparison}
\end{center}
\end{figure*}

There is no unique procedure for deblending objects and defining their object-sky boundaries on astronomical images. Two different approaches can be adopted to tackle this problem, and these two approaches are exemplified by the SDSS {\it PHOTO} pipeline and the SExtractor software \citep{bertin96}. Both approaches are shown in Figure~\ref{deblending-comparison}. In the case of SDSS deblending (Section~\ref{sdss-deblend}), the pipeline tries to isolate the flux in a given object to create a new image of what this object would look like if only this flux was shown. The object area of interest is then determined from this new deblended image. It is important to note that this object area may extend to areas that used to be occupied by neighboring galaxies, and may affect the photometry if some flux from the neighboring galaxies that were removed is left behind. SExtractor deblending (Section~\ref{sex-deblend}) does not attempt to create a new image showing the deblended object. It uses a multi-threshold method to look at the flux tree  of the object in its original image. As it moves up the tree going from one isophotal level to the next, it looks at the flux in each branch. If the flux in a branch is above a set minimum fraction of the total flux in the tree, then this branch is deemed to be a separate object altogether. Once all the objects have been identified, SExtractor then produces a segmentation image that assigns pixels to the different objects. In the case of a galaxy with a close neighbor, the net result is a ``ridge" of segmentation values at the saddle point between the two adjacent objects. The segmentation of an object never extends over the area occupied by its neighbors (see upper left-hand panel of Figure~\ref{deblending-comparison}). It is clear from Figure~\ref{deblending-comparison} that each object segmentation area misses a significant fraction of the flux in the object because it spills over into the segmentation area of its neighbor. The SExtractor magnitudes measured from these areas would therefore systematically underestimate the brightnesses of the two objects. However, if we also fit some surface brightness model (e.g., bulge+disk) to the flux inside a given object segmentation area and then integrate the flux in this model out to large radii, then the missing flux is recovered. As we will show later, such combination of SExtractor deblending {\it and} model photometry can produce more reliable magnitudes and colors in crowded environments.

\subsubsection{SDSS Deblending}\label{sdss-deblend}
 The atlas images provide object fluxes in pixels
deemed to belong to the object by the SDSS deblending algorithm \citep{lupton01},
and they have zero object flux everywhere else. The SDSS pipeline
adds a constant ``softbias'' level of 1000 counts to all pixels in
the atlas and PSF images. This softbias level was subtracted from the
images, so the atlas image sky pixels therefore have a value of
zero. However, these atlas images cannot be used as is for our morphological
analysis. The isophotal cut chosen to define an object is too high,
and the wings of galaxy profiles where useful information can still be
found have been truncated from these images as a result of this
choice. Our science thumbnail images were therefore produced by
extending the object pixel area in the atlas images using pixels from
the corrected image of the field in which the object resides. We ran
SExtractor on the corrected image to identify sky
pixels. Pixels from the corrected image were added to the science
thumbnail only if they were identified as sky pixels.

The mask image is a critical element of the analysis as it specifies
the object pixel brightness distribution that will be ``seen" by the
fitting algorithm. Mask pixels could take on three different values: 1
(object), 0 (sky) and $-$2 (excluded from fit). Initially, our main consideration
was to remain consistent with the object deblending performed by the
SDSS pipeline. We did not perform our own object deblending at first. We
used the atlas image itself as a starting point for the
creation of the mask image. For pixels in the mask image overlapping
with the atlas image, we assigned them a value of 1 when the atlas
image pixel value was greater than 1000 and a value of $-$2
otherwise. We then used the SExtractor segmentation image of the
corrected image to complete the mask. Pixels in the final mask were
assigned a value of 1 if atlas mask pixel value was 1 and a value of 0
if atlas mask was $-$2 and SExtractor mask value was 0. The net result
was a mask with the same dimensions as the atlas image but with more sky pixels than the original atlas images that preserves the object deblending of the SDSS pipeline.

\subsubsection{SExtractor Deblending}\label{sex-deblend}
The procedure for preparing GIM2D postage stamp images using SExtractor rather than SDSS deblending is considerably simpler. It starts directly from the SDSS corrected image frame on which the object is located. SExtractor was run on this corrected frame to measure the parameters X\_IMAGE, Y\_IMAGE, BACKGROUND, and ISOAREA\_IMAGE using a value of 0.00005 for the SExtractor deblending contrast parameter DEBLEND\_MINCONT and a 1-$\sigma$ isophote for the analysis area. The area extracted around each object for the GIM2D postage stamp images was set to five times the area given by ISOAREA\_IMAGE. The GIM2D decompositions were performed over all pixels flagged as object {\it or} background in the SExtractor segmentation image. Objects in the segmentation images of the SDSS corrected frames are sharply delineated by the location of the isophote corresponding to the detection threshold
because SExtractor considers all pixels below this threshold to be background pixels. However, precious information
on the outer parts of the galaxy profile may be contained in
the pixels below that threshold, and fits should therefore not be
restricted only to object pixels to avoid throwing that information
away (This is an analogous problem to the sharp isophotal limit in SDSS atlas images). Pixels belonging to objects in the neighborhood of
the primary object being fit are masked out of the fitting area
using the SExtractor segmentation image. As noted earlier, the flux from the
primary object that would have been in those masked areas in
the absence of neighbors is nonetheless properly included in
the magnitude measurements because GIM2D magnitudes
were obtained by integrating the best-fit models over all pixels.

\subsection{Sky Background Level Determination}\label{skybkg}

Systematic errors in the sky background level determination dominate the errors on the measured structural parameters. An erroneous sky level can also mislead the bulge+disk decomposition algorithm into introducing unphysical bulge or disk components. For example, a very large disk component would result from underestimating the sky level because the positive sky offset would look like such a component to the algorithm. It is therefore critical to measure the best sky possible.

We initially adopted the SDSS sky background levels for our bulge+disk decompositions for the sake of consistency. We used the sky levels given by the keyword SKY in the headers of the corrected images. The sky level for a given corrected frame was subtracted from all the GIM2D science postage stamp images of the objects extracted from this frame. The sky level was then fixed to zero for the bulge+disk decompositions. As discussed later, this procedure did not produce sky background levels that were good enough for our decompositions.

We then used sky background levels and standard deviations determined by GIM2D. GIM2D first uses all the pixels in the science thumbnail
image flagged as background pixels (flag value of zero)
in the SExtractor segmentation image. GIM2D further prunes
this sample of background pixels by excluding any background
pixel that is closer than 4\arcsecpoint0 from {\it any} (primary or neighboring)
object pixels. This buffer zone ensures that the flux from all
SExtracted objects in the image below all the 1.0-$\sigma_{bkg}$ isophotes
does not significantly bias the mean background level upwards
and artificially inflate $\sigma_{bkg}$. A minimum of 20,000 sky pixels
was imposed on the area of the sky region. In cases where
the number of sky pixels in the input science thumbnail image
was insufficient, the original SDSS corrected image was searched for
the 20,000 sky pixels nearest to the object. Background parameters were re-calculated with GIM2D before
fitting, and the residual background levels were then frozen
to their recalculated values for the bulge+disk fits.

\subsection{Two-Dimensional Bulge+Disk Decompositions}\label{bdfit}

Galaxy structural parameters were measured from bulge+disk
decompositions performed using version 3.2 of the GIM2D software package
\citep{simard02}. We used the sum of a pure exponential disk and a de
Vaucouleurs bulge (S\'ersic index $n_b = 4$) as our galaxy image
model. The free fitting parameters of this model were the total flux
$F$ in data units (DU), the bulge fraction $B/T$ ($\equiv$ 0 for pure
disk systems), the bulge semi-major axis effective radius $r_e$, the
bulge ellipticity $e$ ($e \equiv 1-b/a$, $b \equiv$ semi-minor axis,
$a \equiv$ semi-major axis), the bulge position angle of the major
axis $\phi_{b}$ on the image (clockwise, y-axis $\equiv$ 0), the
semi-major axis exponential scale length $r_d$ (also denoted $h$ in
the literature), the disk inclination $i$ (face-on $\equiv$ 0), the
disk position angle $\phi_d$ on the image, and the $dx$ and $dy$ offsets
of the model center with respect to the SDSS object position on the sky. The background residual level $db$ and the bulge S\'ersic index $n_b$
were held fixed for the fits. The position
angles $\phi_b$ and $\phi_d$ were not forced to be equal for two
reasons: (1) a large difference between these position angles is often a
signature of strongly barred galaxies, and (2) some observed galaxies do have
{\it bona fide} bulges that are not quite aligned with the disk
position angle. In addition to bulge+disk structural parameters, GIM2D
computes a variety of asymmetry indices that can be used to quantify
galaxy substructure. The impact of substructures on GIM2D measurements has been extensively simulated and discussed in \citet{simard02}. We {\it specifically} chose to use a bulge+disk model rather than a more complicated combination (e.g., bulge+disk+bar(+others)) for the sake of keeping our SDSS measurements consistent with what is achievable with galaxy structural measurements at high-redshift.  It is already a challenge to perform reliable bulge+disk decompositions of high redshift galaxy images even with the spatial resolution of the Hubble Space Telescope, and it is not currently possible to use structural models with more components without compromising convergence and avoiding parameter degeneracies.

\begin{figure*}
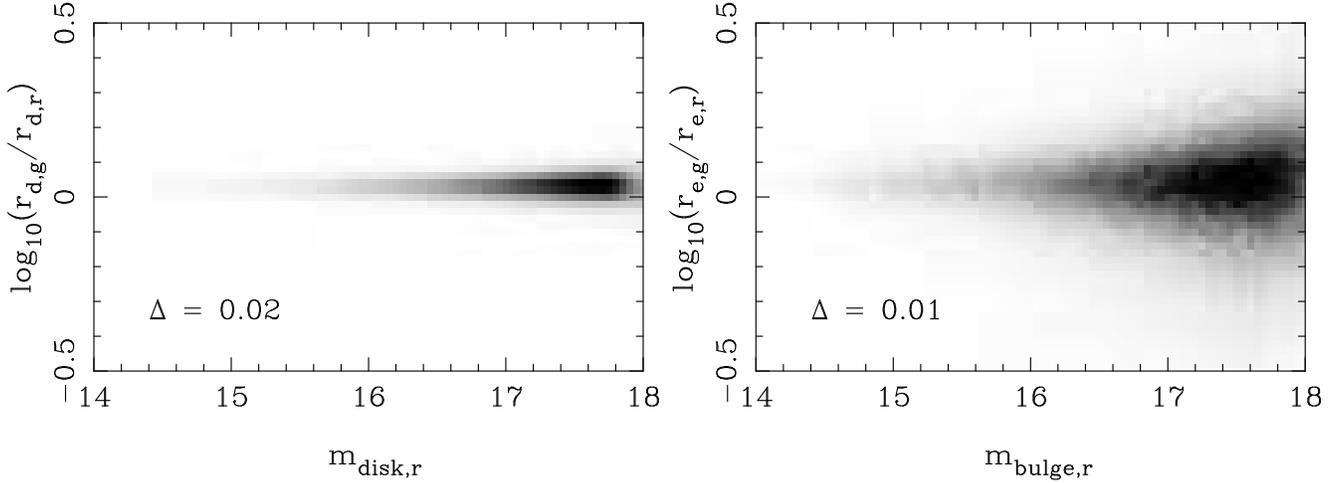

\includegraphics[angle=270,width=8.7cm]{fig4a.eps}
\includegraphics[angle=270,width=8.7cm]{fig4b.eps}
\caption{{\it Left-hand panel}: Log ratio of $g$- to $r$-band disk scale lengths measured from separate bandpass fits as a function of disk magnitude for galaxies with $r$-band bulge fraction $(B/T)_r \leq$ 0.5 and apparent half-light radius $\geq$ 2\arcsecpoint 0. The average ratio of the two disk scale lengths is 10$^{0.02}$ = 1.05. {\it Right-hand panel}: Log ratio of $g$- to $r$-band bulge effective radii measured from separate bandpass fits as a function of bulge  magnitude for galaxies with $r$-band bulge fraction $(B/T)_r > $ 0.5 and apparent half-light radius $\geq$ 2\arcsecpoint 0. The average ratio of the two bulge effective radii is 10$^{0.01}$ = 1.02.}
\label{rdg_rdr}
\end{figure*}

Fits in the $g$- and $r$- bands were done using the two-bandpass separate and simultaneous fitting
procedures described in \citet{simard02}. For the separate, two-bandpass fits, the $g$- and $r$-band images were fitted independently. For the simultaneous, two-bandpass fits, the images were fitted by forcing the bulge radius, ellipticity and position angle and disk scale length, inclination and position angle  to take on the same values in both bandpasses. The main advantage of the simultaneous fits is to minimize the errors on the measured structural parameters because all of the data is used at once. Their main disadvantage is that they are blind to color gradients, and significant color gradients would make this approach invalid because sizes would be significantly different in different bandpasses. To test the validity of the application of simultaneous fitting to the SDSS data, we plotted the ratio of the $g$-band and $r$-band disk scale lengths from separate fits as a function of disk magnitude (Figure~\ref{rdg_rdr}). The two disk scale lengths are within a few percent of one another and in agreement with a recent cosmologically based disk formation model \citep{dutton11a}. As also shown in Figure~\ref{rdg_rdr}, the bulge effective radii measured separately in the two bandpasses agree to 2$\%$. We therefore deem the simultaneous fitting approach to be valid. There is one important difference between the simultaneous fitting procedure of  \citet{simard02}~and the one used here. \citet{simard02} used the same $dx$ and $dy$ offsets in both bands because their images had been registered to the same pixel grid in both bands. This is not the case with the SDSS images, and using the same set of offsets produced ``positive-negative" residuals in the cores of many galaxies. We therefore let $dx$ and $dy$ be different in $g$ and $r$, i.e, we let $(dx)_g$, $(dy)_r$, $(dx)_r$ and $(dy)_r$ be free parameters in the simultaneous fits. The simultaneous fits therefore used a  total number of fourteen free parameters ($F_g$, $F_r$, $(B/T)_g$, $(B/T)_r$, $r_e$, $e$, $\phi_{b}$, $r_d$, $i$, $\phi_{d}$, $(dx)_g$, $(dx)_r$, $(dy)_g$ and $(dy)_r$).

\begin{figure*}
\begin{center}
\includegraphics[angle=0,width=12cm]{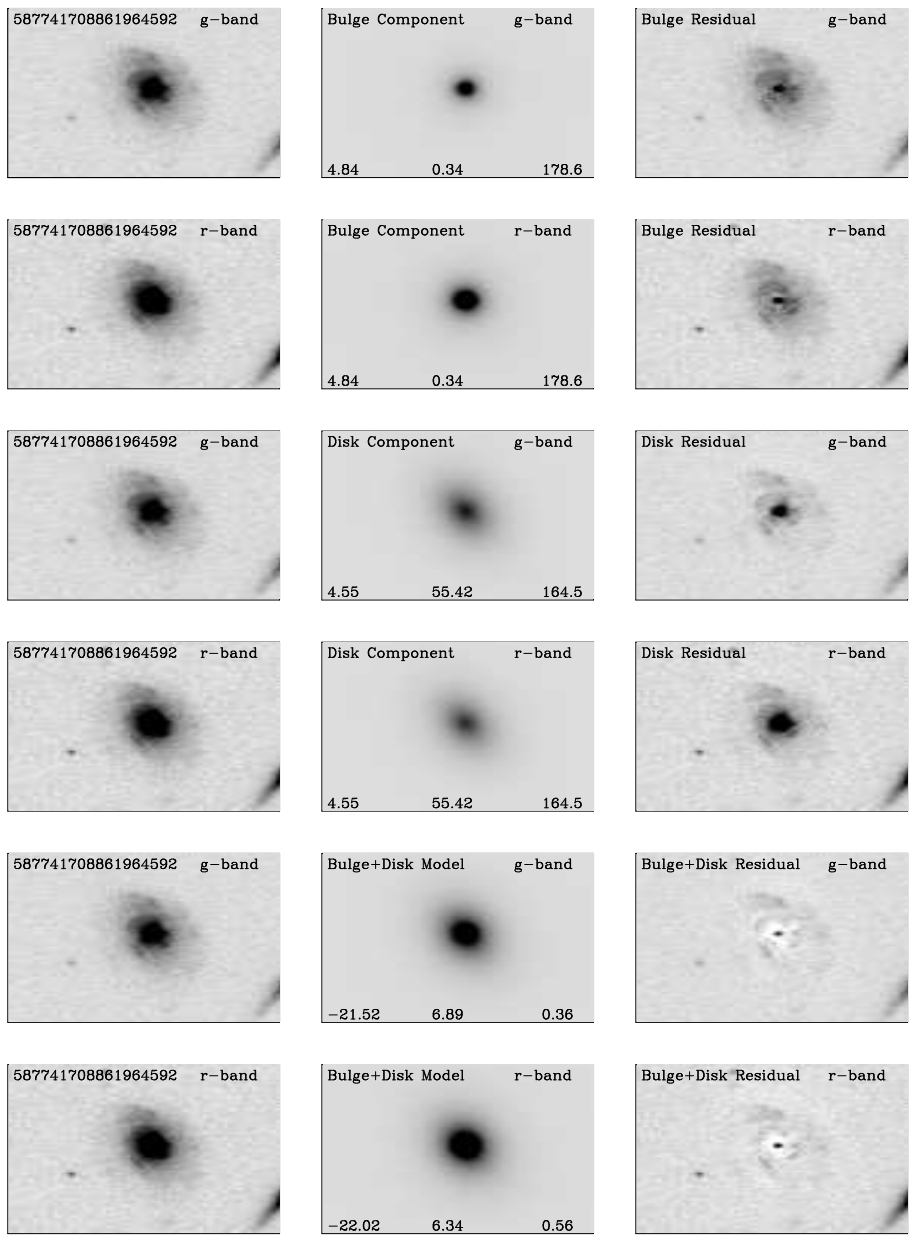}
\caption{Example of a GIM2D bulge+disk decomposition (SDSS objID = 587741708861964592) . From left to right: Galaxy cutout from SDSS corrected image, GIM2D output model image and GIM2D residual image. The top two rows show the bulge component only, the middle two rows show the disk component only, and the bottom two rows show the full bulge+disk model. The labels at the bottom of the bulge component images are effective radius in kiloparsecs, ellipticity ($\equiv$ 0 for circular) and position angle (measured clockwise from $+$y axis. The labels at the bottom of the disk component images are disk scale length in kiloparsecs, inclination angle ($\equiv$ 0 for face-on) and position angle (measured clockwise from $+$y axis). The labels are the bottom of the bulge+disk model images are rest-frame absolute magnitude, radius in kiloparsecs and bulge fraction.  The minimum and maximum values for the greyscale stretch are $<$bkg$> -5\sigma_{\rm bkg}$ and $<$bkg$> +30\sigma_{\rm bkg}$ respectively where $<$bkg$>$ and $\sigma_{\rm bkg}$ are the mean and dispersion of the background pixel values.}
\label{gim2d-fit-example}
\end{center}
\end{figure*}

\begin{figure*}
\begin{center}
\includegraphics[angle=0,width=15cm]{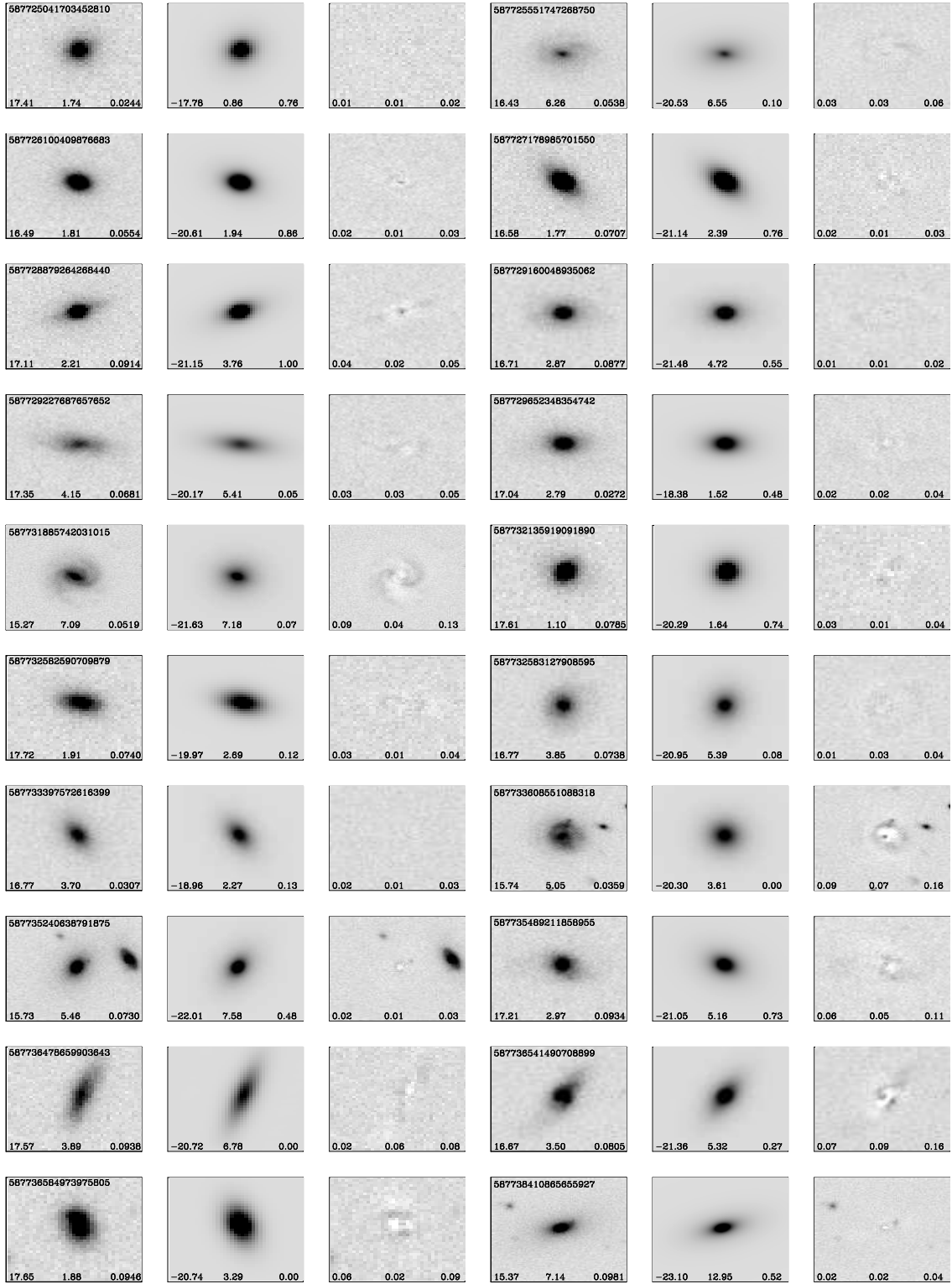}
\caption{Mosaic of $r$-band GIM2D bulge+disk decompositions. Three postage stamp images are shown for each galaxy: galaxy cutout from SDSS corrected image, galaxy bulge+disk GIM2D model convolved with its point-spread-function, and the GIM2D residual image. Each corrected image cutout is labelled by objID, apparent GIM2D model magnitude, GIM2D model half-light radius in arcsecs, and its SDSS redshift. Each GIM2D model image is labelled by GIM2D model rest-frame absolute magnitude, GIM2D model half-light radius in kiloparcsecs and GIM2D model bulge fraction. Each GIM2D residual image is labelled by three asymmetry indices: $RT1\_2$, $RA1\_2$ and $S2 = RT1\_2 + RA1\_2$ (as defined in \citet{simard02,simard09}). The minimum and maximum values for the greyscale stretch for a given galaxy are $<$bkg$> -5\sigma_{\rm bkg}$ and $<$bkg$> +30\sigma_{\rm bkg}$ respectively where $<$bkg$>$ and $\sigma_{\rm bkg}$ are the mean and dispersion of the background pixel values.}
\label{gim2d-fit-mosaic}
\end{center}
\end{figure*}

Four sets of  GIM2D fits were performed on the spectroscopic sample defined in Section~\ref{data} with different combinations of object deblending, sky level determinations and fitting procedures: (1) Separate fitting + SDSS object deblending + SDSS sky level (``SEP+SDSSDEBL+SDSSBKG"), (2) Separate fitting + SDSS deblending + GIM2D sky level (``SEP+SDSSDEBL+GM2DSKY"), (3) Simultaneous fitting + SDSS deblending + GIM2D sky level (``SIM+SDSSDEBL+GM2DSKY"),  and (4) Simultaneous fitting + SExtractor deblending + GIM2D sky level (``SIM+SEXTDEBL+GM2DSKY"). This last set of fits was adopted as our preferred set on the basis of quality assessment tests described in Section~\ref{QAmetrics}, and the procedure was then run on our full photometric sample of 1.12 million galaxies. Out of all the galaxies in this sample, required files could not be obtained from the SDSS image server for 58 galaxies, and GIM2D convergence was not achieved for 7361 objects (failure rate of 0.66$\%$). Visual inspection of these failures traced most of them back to artefacts (e.g., image defects, bright star spikes, bits of a ``shredded" galaxy) in the SDSS catalog. Figure~\ref{gim2d-fit-example} shows a detailed example of a bulge+disk decomposition of a single galaxy, and the summary decompositions for twenty galaxies drawn at random from our sample are shown in Figure~\ref{gim2d-fit-mosaic}.

In addition to our canonical $n_b = 4$ bulge + disk fitting model (denoted as ``n4" hereafter), we used two other fitting models for comparison. The first additional model was a free $n_b$ bulge + disk model (denoted as ``fn" hereafter) with $n_b$ allowed to vary from 0.5 to 8, and the second one was a single component, pure S\'ersic model (denoted as ``pS" hereafter). The fitting parameters of the latter were $F_g$, $F_r$, $r_e$, $e$, $\phi_{b}$,  $(dx)_g$, $(dx)_r$, $(dy)_g$, $(dy)_r$ and $n_g$ where $n_g$ is now the global galaxy S\'ersic index, and $n_g$ was also allowed to vary from 0.5 to 8. All of these additional fits were performed with the same object deblending and sky background level as the ``SIM+SEXTDEBL+GM2DSKY" dataset described above. Structural parameters from all three fitting models are compared in Section~\ref{cmpmodels} and given in the data tables in Section~\ref{bigtables}. 

All of our galaxy model fluxes were converted to SDSS magnitudes
using:

\begin{eqnarray}
m_{galaxy} = -2.5 {\rm log_{10}} (F/t) - \chi {\rm sec} z - m_0
\label{g2dmag}
\end{eqnarray}
\noindent where $F$ is the model flux, $t$ is the exposure time of the SDSS corrected images
(53.907456 seconds), sec $z$ is the airmass, $\chi$ is the extinction
coefficient, and $m_0$ is the magnitude zeropoint. Values for these
coefficients were taken directly from the SDSS database table {\tt Field} and
applied on a field-by-field basis. The photometric coefficients from this table put our flux measurements on the SDSS $ugriz$ magnitude system. We did not transform our magnitudes from the SDSS to the AB system. The largest offset between the two systems is found in the $u-$band where $u_{AB}$ = $u_{SDSS} - 0.04$. The offsets in the other bands are all 0.01 mag or less. 

\subsection{Rest-Frame Quantities}\label{rest-frame}

All angular sizes were converted to physical
sizes according to the equation:

\begin{eqnarray}
R = r \frac{c}{1000H_0} \frac{1}{1+z}\int_{0}^{z} \frac{dz^{\prime}}{\sqrt{\Omega_m (1+z^{\prime})^3 + \Omega_\Lambda}}
\label{ang-diam}
\end{eqnarray}

\noindent where $z$ is the redshift, $c$ is the speed of light, $r$
is the measured angular size in radians, and $R$ is the
corresponding physical size in kiloparsecs.
Equation~\ref{ang-diam} is only valid for flat ($\Omega_k \equiv 1 - \Omega_m - \Omega_\Lambda = 0$) cosmologies.

The $g$-band and $r$-band galaxy, bulge and disk rest-frame absolute magnitudes were computed from the GIM2D apparent magnitudes
according to the equations:

\begin{subequations}
\begin{eqnarray}
	  M_{g,g} & = & m_{g,g} - e_g - DM(z) -k_g
	  \label{gmag-galaxy}
	  \\
          M_{r,g} & = & m_{r,g} - e_r - DM(z) - k_r
	  \label{rmag-galaxy}
	  \\
	  M_{g,b} & = & m_{g,g} - e_g - 2.5 {\rm log_{10}} ((B/T)_g) \nonumber
	  \\ 
	  & & - DM(z) - k_g
	  \label{gmag-bulge}
	  \\
          M_{r,b} & = & m_{r,g} - e_r - 2.5 {\rm log_{10}} ((B/T)_r) \nonumber 
	  \\ 
	  & & - DM(z) - k_r
	  \label{rmag-bulge}
	  \\
	  M_{g,d} & = & m_{g,g} - e_g - 2.5 {\rm log_{10}} (1-(B/T)_g) \nonumber
	  \\ 
	  & & - DM(z) - k_g
	  \label{gmag-disk}
	  \\
          M_{r,d} & = & m_{r,g} - e_r - 2.5 {\rm log_{10}} (1-(B/T)_r) \nonumber 
	  \\ 
	  & & - DM(z) - k_r
          \label{rmag-disk}
\end{eqnarray}
\end{subequations}

\noindent where $m_{g,g}$ and $m_{r,g}$ are the $g$-band and $r$-band observed GIM2D galaxy
model magnitudes from Equation~\ref{g2dmag}, $e_g$ and $e_r$ are the line-of-sight galactic
extinctions in magnitude from the SDSS database, $(B/T)_g$ and
$(B/T)_r$ are the GIM2D observed bulge fractions, $DM(z)$ is the distance modulus for redshift $z$,
and $k_g$ and $k_r$ are the $k$-corrections.  The $k$-corrections were computed using {\tt kcorrect} version 4 \citep{blanton07}. Rest-frame magnitudes are included in the data table presented in Section~\ref{bigtables}.

\subsection{Sample Volume Corrections}\label{vmax-corr}

Volume corrections must be applied to galaxies in our sample to properly account for the effects of selection on the visibility of different classes of galaxies. The SDSS spectroscopic sample is complete down to a $r$-band magnitude of 17.77 and an effective surface brightness $\mu_{50,r}$ of 23 mag arcsec$^{-2}$ (corrected for Galactic extinction). Our  $V_{max}$ corrections are similar to the ones in \citet{shen03}. First, the magnitude range $r_{min} \leq r \leq r_{max}$ corresponds to a maximum redshift $z_{max,m}$ and a minimum redshift $z_{min,m}$:

\begin{subequations}
\begin{eqnarray}
d_L(z_{max,m}) & = & d_L(z) 10^{-0.2(r-r_{max})}
\label{vmax-dlmax}
\\
d_L(z_{min,m}) & = & d_L(z) 10^{-0.2(r-r_{min})}
\label{vmax-dlmin}
\end{eqnarray}
\end{subequations}

\noindent where $d_L(z)$ is the luminosity distance at redshift $z$. The surface brightness limit constrains $V_{max}$ for a given galaxy mainly through  the $(1+z)^4$ dimming effect. The maximum redshift at which a galaxy of surface brightness $\mu_{50,r}$ at $z$ can still be detected within a limiting surface brightness $\mu_{lim}$ = 23 is given by:

\begin{eqnarray}
z_{max,\mu} & = & (1 + z)10^{(23.0-\mu_{50,r})/10} - 1
\label{vmax-mu}
\end{eqnarray}

One difference between the $V_{max}$ corrections in \citet{shen03} and the ones used here is that we do not impose a minimum size limit to galaxies in our sample. 
The real maximum and minimum redshifts, $z_{max}$ and $z_{min}$, for a given galaxy are therefore:

\begin{subequations}
\begin{eqnarray}
z_{min} & = & {\rm max}(z_{min,m}, 0.005)
\label{vmax-zmin}
\\
z_{max} & = & {\rm min}(z_{max,m}, z_{max,\mu})
\label{vmax-zmax}
\end{eqnarray}
\end{subequations}

With the above limits, $V_{max}$ is then given by the integral:

\begin{eqnarray}
V_{max} = \frac{1}{4\pi} \int d\Omega f(\theta, \phi) \int^{z_{max}(\theta,\phi)}_{z_{min}(\theta,\phi)} \frac{d^2_A(z)}{H(z)(1+z)} c dz
\label{vmax-integral}
\end{eqnarray}

\noindent where $H(z)$ is the Hubble parameter at redshift $z$, $c$ is the speed of light, $f(\theta,\phi)$ is the sampling fraction as a function of position on the sky, and $\Omega$ is the solid angle. We take the function $f(\theta, \phi)$ to be a constant over the entire 8032 square degree areal coverage of the SDSS DR7 Legacy survey. The angular part of Equation~\ref{vmax-integral} thus becomes $(1/4\pi) (8032/41253) (4\pi)$ = 0.1947. $V_{max}$ values are given in the data table presented in Section~\ref{bigtables}.

\section{Results}\label{results}

\subsection{Quality Assessment}\label{QAmetrics}

We based the quality assessment of our measured structural parameters on three science-motivated metrics.

\subsubsection{Size-Luminosity Relation of Disks}\label{lumsize}

\begin{figure}
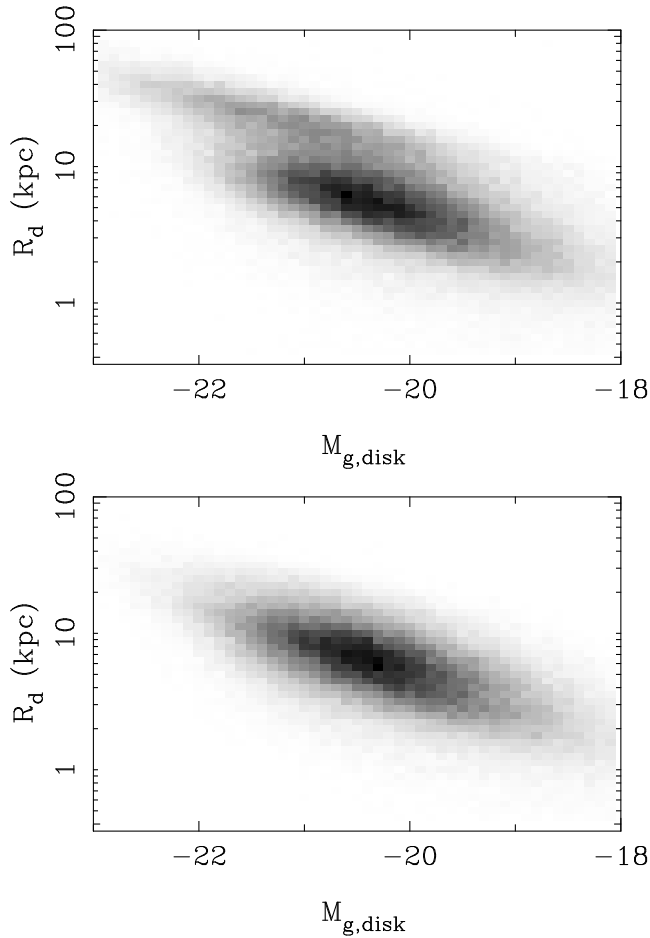

\includegraphics[angle=270,width=8.5cm]{fig7a.eps}
\hfill
\includegraphics[angle=270,width=8.5cm]{fig7b.eps}
\caption{{\it Top panel}: $g$-band disk size-luminosity relation without volume corrections for face-on (0.75 $\leq$ b/a $\leq$ 1.0) disks from separate fits with SDSS background levels (``SEP+SDSSDEBL+SDSSBKG"). {\it Bottom panel}: Same disk sample but with GIM2D-determined sky background levels (``SEP+SDSSDEBL+GM2DBKG"). Note the disappearance of the ``large disks" branch with the use of GIM2D sky levels.}
\label{disk-lumsize}
\end{figure}

Figure~\ref{disk-lumsize} shows the size-luminosity of disks in the $g$-band. Disks are expected to follow a well-defined and well-known scaling relation, and most of the disks do follow such a relation. However, some disks in the ``SEP+SDSSDEBL+SDSSBKG" photometric dataset lie on an ``upper branch" of large disks that appear to be 2-3 times larger at a given disk luminosity. Plotting the same relation for the $u$-band showed  an increasing bimodality in the sizes of disks. Visual inspection of these apparently large disks showed that they were in fact caused by constant sky background residuals left after subtracting the SDSS sky values. As explained in Section~\ref{skybkg}, our initial procedure assumed that the SDSS sky values were ``perfect" and that the GIM2D fits could be performed by fixing the sky value to zero once the SDSS sky values had been subtracted from the input science images. These constant positive sky residuals were  being fit by the algorithm with a very large (and thus very flat) disk component. The false disks lie on a linear relation because of the strong covariance between luminosity and size errors through the relation $F = 2\pi\Sigma_0 r_d$ for the total flux $F$ of an exponential disk with central surface brightness $\Sigma_0$ and scale length $r_d$. Recomputing local sky background levels following the procedure described in Section~\ref{skybkg} corrected this problem as shown by the ``SEP+SDSSDEBL+GM2DBKG" photometric dataset in the bottom panel of  Figure~\ref{disk-lumsize}. The $u$-band relation was also corrected by the same procedure. The disk size-luminosity relation is therefore a sensitive test of our sky background level determinations.

It is worth noting here that this sky background problem is different from the well-known problem with the photometry of bright galaxies in SDSS \citep[][and references therein]{abazajian09}. The SDSS photometric pipeline produces systematic errors in the estimation of the sky near bright ($r < 16$) galaxies that cause their fluxes and scale sizes to be underestimated. The sky background problem encountered here was related to positive sky residuals.  These effects of these residuals became especially significant here because sky level errors dominate the systematic errors in bulge+disk decompositions \citep{simard02}. It will also be more significant in more crowded environments (including close pairs of galaxies).

\subsubsection{Integrated Galaxy Magnitudes and Colors}\label{cmds}

\begin{figure*}
\begin{center}
\resizebox{\textwidth}{!}{\rotatebox{-90}{\includegraphics{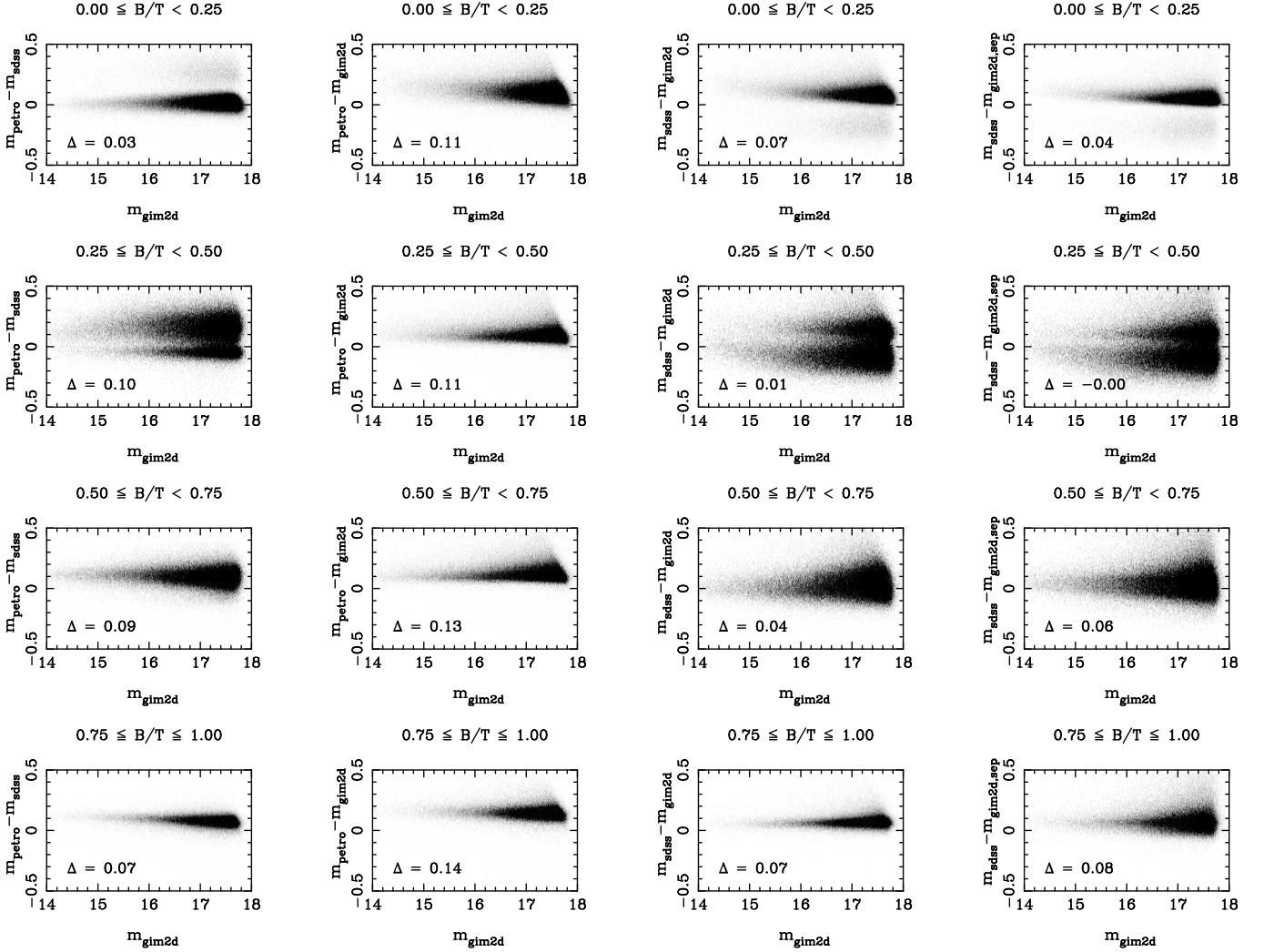}}}
\caption{Comparison of $r$-band GIM2D magnitudes with Petrosian and model SDSS magnitudes (``modelMag") as a function of $r$-band bulge fraction.  {\it First column}: SDSS Petrosian versus SDSS model. {\it Second column}: SDSS Petrosian versus GIM2D (``SIM+SEXTDBL+GM2DBKG"). {\it Third column}: SDSS model versus GIM2D (``SIM+SEXTDBL+GM2DBKG"). {\it Fourth column}: SDSS model versus GIM2D (``SEP+SDSSDBL+SDSSBKG" denoted $m_{gim2d,sep}$ here). Note that the panels in the first column do not depend on any GIM2D photometry. }
\label{dr7_mags_cmp}
\end{center}
\end{figure*}

Figure~\ref{dr7_mags_cmp} compares our  GIM2D-based total magnitudes with the SDSS Petrosian and model magnitudes as a function of bulge fraction.  There are two different types of SDSS model magnitudes. The first type (called ``modelMag") comes from fitting the object with a pure deVaucouleurs and a pure exponential profile, determining which profile fits the object better and computing the magnitude of this profile. The  ``modelMag" magnitudes are widely used throughout the literature because they are the ones explicitly given in the SDSS database. The second type of model magnitudes (called ``cmodel") is computed from the composite flux given by the linear combination of the two profiles, i.e., $F_{composite} = (fracDeV)F_{deV}+ (1-fracDeV)F_{exp}$ (the value of {\it fracDev} is stored in the SDSS database). Although it would be more fair to compare the GIM2D magnitudes to the ``cmodel" magnitudes, we decided to compare our photometry with the most widely used set of SDSS model magnitudes to emphasize that the use of SDSS model magnitudes requires some caution for galaxies of intermediate morphological types. This is obvious from looking at the panels in the first column of Figure~\ref{dr7_mags_cmp}. These panels do not depend on any GIM2D photometry. They show a ``bifurcation" at intermediate bulge fraction values because the ``modelMag" model magnitudes can only come from either a pure disk or a pure bulge model, and some objects with a mixture of bulge and disk components become ambiguous enough to such a binary classification that they end up in different bins. 

The second and third columns of Figure~\ref{dr7_mags_cmp} show that the offset between GIM2D and SDSS magnitudes is independent of bulge fraction. This lack of dependence on the type of galaxy light profiles makes it easier to compare the different sets of magnitudes. \citet{blanton03} reported a systematic trend between their S\'ersic-based magnitude and SDSS Petrosian magnitudes as a function of the S\'ersic index value with Petrosian magnitudes underestimating the $g$- and $r$-band fluxes by 20$\%$ in their high ($n \simeq$ 4) S\'ersic index galaxies. The bifurcation at intermediate bulge fractions is also visible in the comparison between GIM2D and SDSS model magnitudes.

\begin{figure*}
\begin{center}
\includegraphics[angle=0,width=15cm]{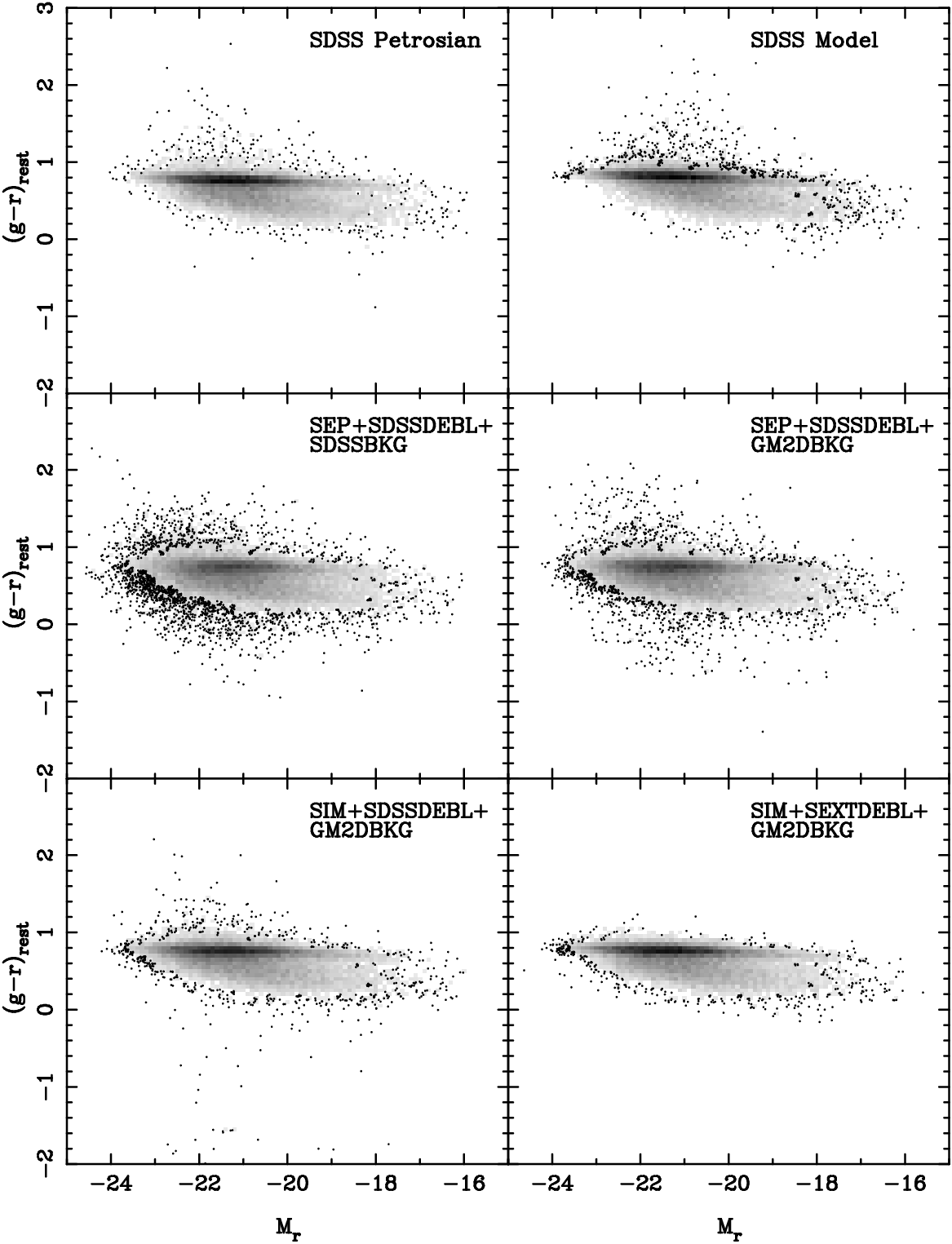}
\caption{Comparison of rest-frame color-magnitude diagrams for the \citet{patton11} DR7 galaxy pair sample  of ($N$=22565): {\it Top left}: SDSS Petrosian magnitudes {\it Top right}: SDSS model magnitudes {\it Middle left}: GIM2D separate fits  with SDSS sky background and deblending, {\it Middle right}: GIM2D separate fits  with GIM2D sky background and SDSS deblending {\it Bottom left}: GIM2D simultaneous fits with GIM2D sky background and SDSS deblending. {\it Bottom right}: GIM2D simultaneous fits with GIM2D sky background and SExtractor deblending. The same greyscale was used in all subpanels. The bin size for the greyscale was ($\Delta(g-r)$, $\Delta M_r$) = (0.1, 0.05). The greyscale was replaced by the actual data points wherever the number of points in a given bin was less than 2.}
\label{dr7_cmds_cmp_pairs}
\end{center}
\end{figure*}

\begin{figure*}
\begin{center}
\includegraphics[angle=0,width=15cm]{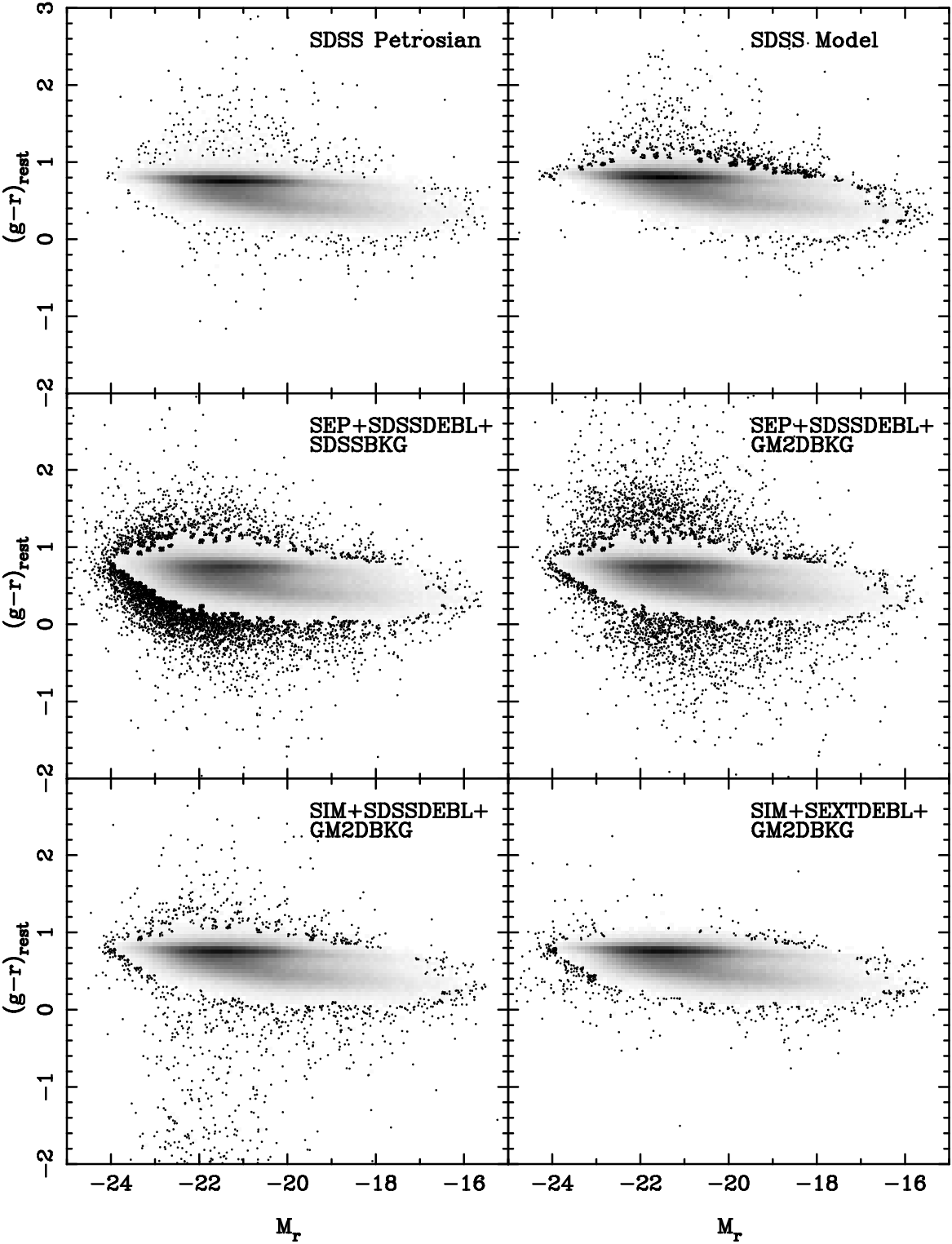}
\caption{Comparison of rest-frame color-magnitude diagrams for the \citet{patton11} DR7 galaxy pair control sample (N=290090): (Top left) SDSS Petrosian magnitudes, (top right) SDSS model magnitudes, (middle left) GIM2D separate fits  with SDSS sky background and deblending, (middle right) GIM2D separate fits  with GIM2D sky background and SDSS deblending, (bottom left) GIM2D simultaneous fits with GIM2D sky background and SDSS deblending, and (bottom right) GIM2D simultaneous fits with GIM2D sky background and SExtractor deblending. The same greyscale was used in all subpanels. The bin size for the greyscale was ($\Delta(g-r)$, $\Delta M_r$) = (0.1, 0.05). The greyscale was replaced by the actual data points wherever the number of points in a given bin was less than 2.}
\label{dr7_cmds_cmp_ctrl}
\end{center}
\end{figure*}

Object deblending can have a significant impact on galaxy photometry especially in crowded environments (Section~\ref{defining-objects}). The simplest case of a crowded environment is a close pair of galaxies, and we tested our GIM2D photometry using the galaxy pair sample of  \citet{patton11}. This galaxy pair sample also comes with a control sample matched in stellar mass and redshift, and we used this control sample for a comparison between crowded and non-crowded environments. The control sample provides a cleaner comparison than the entire SDSS sample in the sense that selection effects due to galaxy properties such as luminosity and morphological type and technical considerations such as fiber collision are the same for both samples. Figures~\ref{dr7_cmds_cmp_pairs} and~\ref{dr7_cmds_cmp_ctrl} show color-magnitude diagrams for the Patton et al. galaxy pair and control samples.  These diagrams are shown for different types of photometry.   The GIM2D separate fits (``SEP+SDSSDEBL+SDSSBKG") produce a red sequence and a blue cloud that are less well-defined than those based on Petrosian and model magnitudes. GIM2D magnitudes are {\it total} magnitudes in the sense that the fluxes in the best-fit bulge and disk components are integrated to infinity, and as such, they will have a lower signal-to-noise ratio than Petrosian magnitudes, say, that aim to minimize the amount of missing flux while optimizing the signal-to-noise ratio of the photometric measurements. The GIM2D fits also yield a large number of bright galaxies over the range $-24 \leq M_r \leq 21$. As discussed in Section~\ref{lumsize}, the GIM2D-determined sky backgrounds yielded better disk sizes and luminosities. Adopting this new background level determination method also decreases the number of those bright outliers, but significant scatter towards bluer and redder colors remains. 

Switching from bulge+disk fits performed separately in each filter to the ones performed simultaneously in both bandpasses immediately improves the integrated magnitudes in two important ways. The number of blue and red outliers is now significantly lower, and the red sequence and blue cloud are now as well-defined as for the SDSS photometry. The simultaneous fits produce smaller errors because they simultaneously make use of all the data available for a given object. Visual inspection of the remaining outliers showed them to be in crowded environments where the SDSS object deblending was failing. Adopting the SExtractor deblending (in addition to the simultaneous fits) described in Section~\ref{sex-deblend} eliminated nearly all of the remaining outliers to produce a very clean color-magnitude diagram. The lack of red outliers in the final GIM2D color-magnitude diagram is especially noteworthy. As explained in \cite{patton11}, the presence of these outliers in the official SDSS photometry was ascribed in the literature to a new population of extremely-red pair galaxies \citep[e.g.,][]{alonso06,perez09,darg10}. The GIM2D photometry given here shows that they are in fact due to crowding errors from incorrect deblending. The crowding errors in SDSS preferentially scatter galaxy colors towards the red for the following reason. The $r$-band images go deeper in surface brightness than the $g$-band ones because the SDSS detectors are more red sensitive. The outer isophotes of two neighboring galaxies will therefore overlap more in the red than in the blue, and the SDSS deblender will leave more faint isophotal light behind in the red. When this residual light for a galaxy is included in the photometry of its neighbor(s), it yields (a) redder neighbor galaxy color(s).

\subsubsection{Fiber Colors}\label{fibercolors}

\begin{figure*}
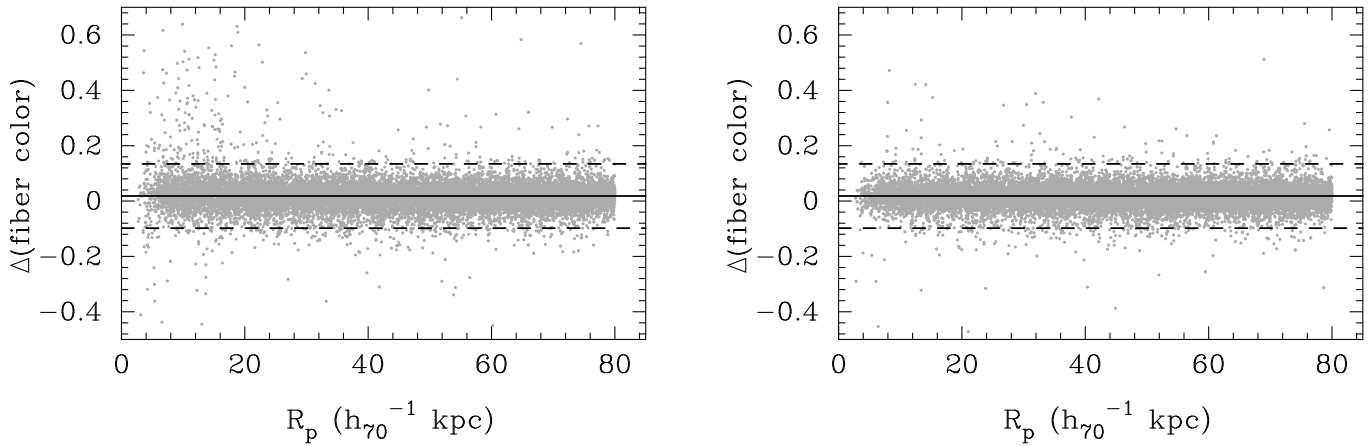

\includegraphics[angle=270,width=8.5cm]{fig11a.eps}
\hfill
\includegraphics[angle=270,width=8.5cm]{fig11b.eps}
\caption{$\Delta$(fiber color) as a function of galaxy pair separation in kiloparsecs  with SDSS deblending ({\it left-hand panel}) and SExtractor deblending ({\it right-hand panel}). $\Delta$(fiber color) is defined as $(g-r)_{gim2d,fiber} - (g-r)_{SDSS,fiber}$.  The lines show the median fiber color (solid line, median value = 0.0182) from SDSS deblending and the 1-$\sigma$ envelope (dashed lines, $\sigma$ = 0.116). The same lines are reproduced in both panels. Note the excess of outliers in the left panel at small separations due to poor deblending.}
\label{delta-fiber-cmp}
\end{figure*}

Our third data quality metric is based on fiber apparent magnitudes and colors. The fiber magnitude of a given object is computed from the flux measured within the aperture of a spectroscopic fiber. The diameter of this aperture is 3$^{\prime\prime}$. The fiber magnitudes are given in the SDSS database.  Given the apparent sizes of the SDSS galaxies with respect to this fiber aperture, fiber magnitudes (and colors) are most often measurements of the central part of a galaxy. The typical fiber covering fraction, which is the ratio of the $g$-band Petrosian to fiber fluxes, is 30$\%$ \citep{ellison08}. We can also compute fiber magnitudes from the GIM2D output model images for comparison with the SDSS fiber quantities. The fiber aperture on the GIM2D model images was centered on the SDSS position of the galaxy on the sky ($(dx,dy)$ = (0,0)) and not on the center of the GIM2D model itself which was allowed to move for the fits. Using the SDSS and GIM2D fiber magnitudes, we then computed a $\Delta$(fiber color) given by $(g-r)_{gim2d,fiber} - (g-r)_{SDSS,fiber}$. This $\Delta$(fiber color) is plotted in Figure~\ref{delta-fiber-cmp} as a function of projected galaxy-galaxy separation on the sky for galaxies in the pair sample of \citet{patton11}. The median $\Delta$(fiber color)  value is not zero due to PSF effects. The SDSS fiber magnitudes were computed by uniformly convolving all the SDSS images to a 2\arcsec seeing whereas the typical size of the SDSS PSFs used for the GIM2D model images is 1\arcsecpoint 4 FWHM. This narrower PSF FWHM spreads less bulge (i.e., red) light outside of the fiber aperture thus making the GIM2D fiber colors slightly redder. A large positive or negative $\Delta$(fiber color) indicates that a bulge+disk decomposition failed because it did not even reproduce the central parts of the galaxy. SDSS deblending produces a large number of outliers, and the number of these outliers increases with decreasing pair separation. The number of galaxies with large $\Delta$(fiber color) values increases significantly at separations less than 20 kiloparsecs which corresponds to angular separations less than 10\arcsecpoint8 at $z = 0.1$.  Switching to SExtractor deblending eliminates the vast majority of the outliers, and no trend is seen anymore as a function of pair separation. This is another example where SExtractor deblending produces more reliable photometry in crowded environments. 

\subsection{Galaxy Fitting Models}\label{cmpmodels}

\subsubsection{Selecting the Appropriate Model with the F-Statistic}\label{ftest}
It is not always true that a galaxy will be appropriately modelled by a multi-component photometric model; in instances where galaxies are best represented by a pure bulge or disk or, alternatively, the data cannot support a more complex model decomposition due to low signal-to-noise or poor spatial sampling, it is unclear how to treat the output of our ``n4" or ``fn" fits.  We therefore want a way to assess the appropriateness of different model decompositions, and a means of deciding when a complex model, e.g. bulge+disk, is to be preferred over a pure S\'ersic profile. We approach this question of finding the appropriate fitting model using the $F$-test to compare the $\chi^2$ residuals of our ``pS", ``n4", and ``fn" fits (see Section~\ref{bdfit}).  The $F$-test provides a quantitative means of judging the relative likelihood of different structural decompositions.  There are of course other approaches \citep[e.g.,][]{allen06}, but the $F$-test seems to be a powerful and straightforward approach as we show below. For each model we estimate the reduced $\chi^2$ using the fit residuals and the number of resolution elements per fit, $n_\mathrm{res}$, in this instance estimated as $n_\mathrm{res} = n_\mathrm{pixels}/(\pi\Theta^2)$, where $n_\mathrm{pixels}$ is the number of unmasked (object) pixels used in the fit and $\Theta$ is the $r$-band seeing half-width at half maximum (HWHM) in pixels. The computed $F$ values for a given pair of models, e.g. ``fn" vs. ``pS", can then be translated into a corresponding probability that a smaller $F$ would be observed under the null hypothesis that a galaxy is similarly well fit by the two decompositions (i.e. that their rms flux residuals are comparable); these probabilities are given in Tables~\ref{sdss_data_table_bd_n4} and~\ref{sdss_data_table_bd_fn}. In a given table, we only provide the probability relative to less complex models, such that the probabilities always correspond to the lower-tail probability of the appropriate $F$-distribution. In other words, $P_{pS}$ is the probability that a bulge+disk model is {\it not} required (compared to a pure S\'ersic model), and $P_{n4}$ is the probability that a free $n_b$ bulge + disk model is {\it not} required (compared to a fixed $n_b=4$ bulge + disk model).

\begin{figure*}
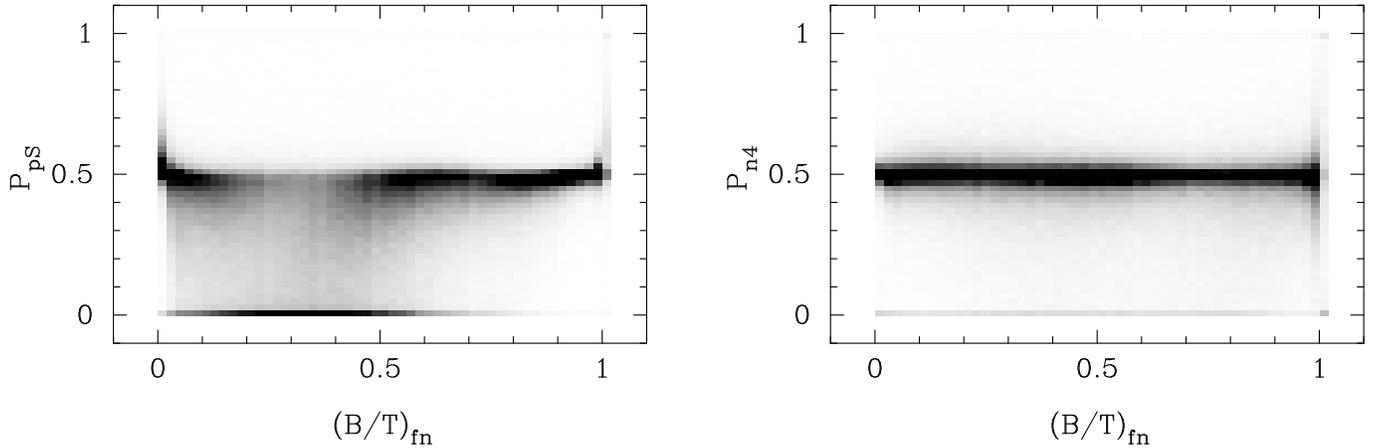

\includegraphics[angle=270,width=8.5cm]{fig12a.eps}
\hfill
\includegraphics[angle=270,width=8.5cm]{fig12b.eps}
\caption{$F$-test probabilities $P_{pS}$ (left panel) and $P_{n4}$ (right panel) versus bulge fraction for the free $n_b$ bulge + disk decompositions. The greyscale represents the two-dimensional distribution of galaxies normalized by the total number of galaxies. The $x$-axis and $y$-axis bin sizes for the two-dimensional distributions were both 0.02. The limits for the greyscale go from zero to 20$\%$ of the peak value of the normalized distribution.}
\label{pps_pn4_btfn}
\end{figure*}

The distributions of $F$-test probabilities versus bulge fraction are shown in Figure~\ref{pps_pn4_btfn}, and they bring forth the limitations of the SDSS imaging in terms of spatial resolution and signal-to-noise. A bulge+disk model is clearly required for galaxies with $0.2 \leq (B/T)_{fn} \leq 0.45$, and galaxies with $(B/T)_{fn}$ $>$ 0.75 do not require a bulge+disk model to fit their light profiles. If we set a value of $P_{pS} \leq 0.32$ as the (1$\sigma$) threshold below which galaxies are likely to be genuine bulge+disk systems, then the fraction of these galaxies in our entire SDSS sample is 26$\%$. The quality of the SDSS imaging is insufficient to determine bulge S\'ersic indices for galaxies in our selected range of apparent magnitudes as $P_{n4}$ versus bulge fraction does not show any statistically significant differences between $n_b=4$ and free $n_b$ models. Only 9$\%$ of the galaxies have $P_{n4} \leq 0.32$. 

\begin{figure*}
\begin{center}
\includegraphics[angle=0,width=15cm]{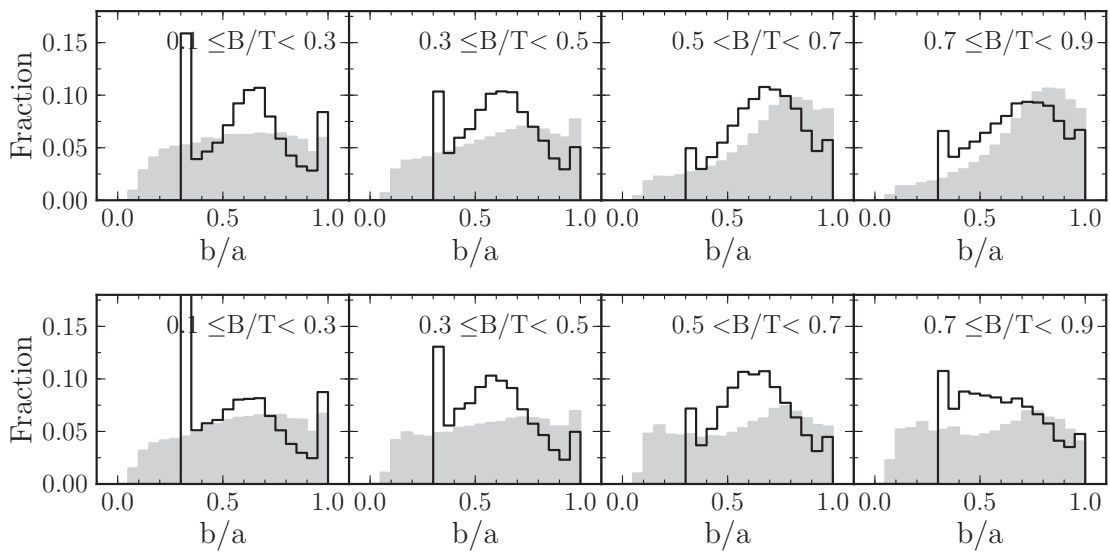}
\caption{Bulge and disk axial ratio distributions for $n_b = 4$ + disk models.  In each panel, shaded and open histograms show the distribution of disk and bulge axial ratios. {\it Top row}: Axial ratio distribution for the full sample of $n_b = 4$ + disk fits.  Note the excess of $b/a > 0.5$ disk components at high $B/T$.  {\it Bottom row}: Axial ratio distribution for the subsample of $n_b = 4$ + disk models where P$_{pS} \leq 0.32$, i.e., where galaxies are likely to be genuine bulge+disk systems.  Note the significant reduction of low-inclination disks, particularly at $B/T > 0.5$.}
\label{ftest_inc.ps}
\end{center}
\end{figure*}

Figure~\ref{ftest_inc.ps} illustrates the usefulness of the $F$-test probabilities to select genuine bulge+disk systems. The axial ratio distribution of a sample of disks randomly inclined in space should be uniform between zero and one modulo some perturbations due to dust and/or bars. Looking at the top row of Figure~\ref{ftest_inc.ps}, one can see that the observed distribution is indeed uniform at low bulge fraction,  but that it also tends towards the same axial ratio distribution as bulges towards higher bulge fractions if no $F$-test selection is made. This behavior would argue that ``disks" in highly bulge-dominated galaxies are not in fact real but were rather introduced by the fitting algorithm as an additional degree of freedom to model the outer wings of a single component galaxy. If we select only galaxies with $P_{pS} \leq 0.32$, then one can see that the resulting disk axial ratio remains uniform even at the highest bulge fractions. We are thus able to select real disks in highly bulge dominated galaxies.

\subsubsection{Structural Parameter Comparison}\label{struct_cmp}

\begin{figure*}
\begin{center}
\resizebox{\textwidth}{!}{\rotatebox{-90}{\includegraphics{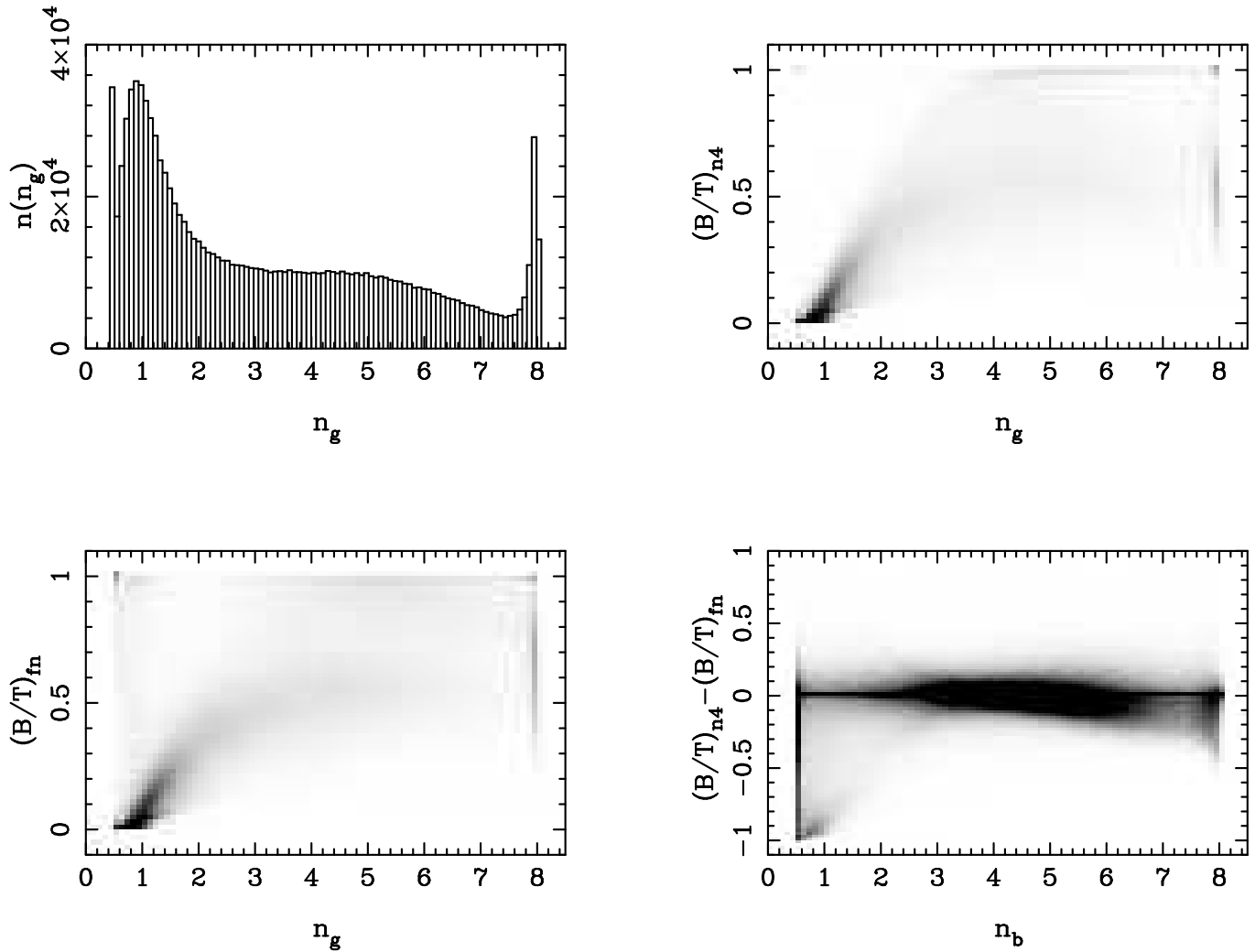}}}
\caption{$r$-band GIM2D galaxy bulge fractions and S\'ersic indices from the three different fitting models. {\it Top left:} Distribution of the galaxy S\'ersic index $n_g$ from the single component, pure S\'ersic fits. {\it Top right:} Bulge fraction $(B/T)_{n4}$ from the $n_b = 4$ bulge + disk decompositions versus galaxy S\'ersic index $n_g$. {\it Bottom left:} Bulge fraction $(B/T)_{fn}$ from the free $n_b$ bulge + disk decompositions versus galaxy S\'ersic index $n_g$. {\it Bottom right:} The difference between $(B/T)_{fn}$ and $(B/T)_{n4}$ as a function of bulge S\'ersic index $n_b$. The greyscale represents the two-dimensional distribution of galaxies normalized by the total number of galaxies. The bin sizes for the two-dimensional distributions were $\Delta(n_g)$ = 0.1 and $\Delta (B/T)$ = 0.02. The limits for the greyscale go from zero to 20$\%$ of the peak value of the normalized distribution.}
\label{dr7_fn_n4_pS_cmp_struct}
\end{center}
\end{figure*}

A detailed comparison between our three different fitting models could easily stand on its own as a separate paper, so we briefly comment here on the salient points of this comparison. We start by looking at measurements of the global galaxy structure from the different models namely the galaxy S\'ersic index $n_g$ and the bulge fraction $B/T$. The distribution of galaxy S\'ersic index values in Figure~\ref{dr7_fn_n4_pS_cmp_struct} shows a peak a sharp peak at $n_g$ = 0.5, a broader and larger peak at $n_g$ = 1, no peak at $n_g$ = 4 and a peak at the maximum allowed S\'ersic value of 8. We will return to the peaks at $n_g$ = 0.5 and $n_g$ = 8 later. The large peak at $n_g$ = 1 reflects the fact that the local galaxy population is dominated by disk galaxies. The lack of a strong peak at $n_g = 4$ is quite interesting because it argues for a lack of global structural similarity in galaxies that are not disk-dominated. In terms of bulge fraction, measured values from both the $n_b = 4$ and free $n_b$ fits (denoted as $(B/T)_{n4}$ and $(B/T)_{fn}$ respectively here) are well correlated with one another and with $n_g$ (Figure~\ref{dr7_fn_n4_pS_cmp_struct}). $(B/T)_{n4}$ and $(B/T)_{fn}$ increase with $n_g$ and reach a value of one at $n_g = 4$. Two ``branches" are seen beyond $n_g = 4$: one at $B/T \sim$ 1 and another one at $B/T \sim 0.5$. The first branch is not surprising, but the second one may point to a potential single- versus double-component degeneracy in the bulge+disk decompositions. In order to understand the behavior of $(B/T)$ at high $n_g$, we looked at the bulge and disk ellipticity difference $\Delta e \equiv e_b - e_d$ versus the ratio of the bulge and disk half-light radii $R_{b/d} \equiv r_e/(1.67r_d)$ in the range $7 \leq n_g <8$ for two ranges in $(B/T)_{n4}$: $0.4 \leq (B/T)_{n4}\leq 0.7$ and $0.9 \leq (B/T)_{n4}< 1.0$. $\Delta e$ is zero which means that the bulges and disks have the same ellipticities i.e., these galaxies may actually have a single component but the fitting algorithm may be using a bulge and disk components to model something that it cannot do even with $(B/T)_{n4}$ = 1 for $n_b$ = 4. The dichotomy actually comes from the ratio of the radii. Galaxies with $0.4 \leq (B/T)_{n4}\leq 0.7$ all have $R_{b/d}$ values of 0.13 with a dispersion of 0.05 whereas galaxies with $0.9 \leq (B/T)_{n4} < 1.0$ have values around 0.56 with a dispersion of 0.16. There are six times more galaxies with $0.4 \leq (B/T)_{n4} \leq 0.7$ than with $0.9 \leq (B/T)_{n4} < 1.0$. The peak in  $R_{b/d}$ where a galaxy ends up seems to depend on the spread in $\Delta e$. The dispersion in $\Delta e$ for galaxies with $0.4 \leq (B/T)_{n4} \leq 0.7$ is twice the dispersion in $\Delta e$ for galaxies with $0.9 \leq (B/T)_{n4} < 1.0$. A smaller $\Delta e$ value for a given galaxy makes it more likely that the fitting algorithm will tend towards a single component model rather than a two-component model because the algorithm will need two components to model a change of ellipticity with radius that does not come from PSF smearing. There is no significant difference between $(B/T)_{n4}$ and $(B/T)_{fn}$ as a function of bulge S\'ersic index except at very low values of $n_b \simeq$ 0.5 where $(B/T)_{n4}$ can be considerably lower than $(B/T)_{fn}$ (Figure~\ref{dr7_fn_n4_pS_cmp_struct}). This behavior at low $n_b$ is expected given that trying to fit a bulge with a very flat profile using a relatively peaky $n_b$=4 component will force the algorithm to minimize the contribution of this $n_b$=4 component as much as possible by converging to a very low bulge fraction. Some previous studies \citep[e.g.,][]{graham08} have reported a dependence of $(B/T)$ on $n_b$. Such a dependence is not seen here, but this may be due to the lack of constraint on bulge profile shape from the SDSS images of the galaxies in our sample as discussed next.

\begin{figure}
\includegraphics[angle=270,width=8.7cm]{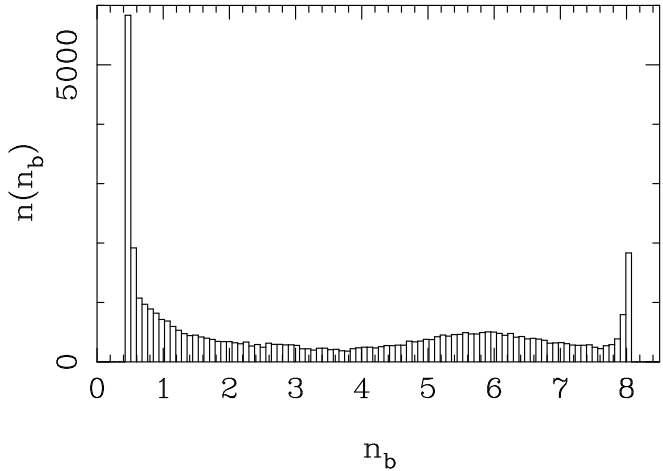}
\caption{Distribution of GIM2D bulge S\'ersic index $n_b$ from the free $n_b$ bulge + disk decompositions for galaxies with $(B/T)_{fn}  > 0$ and $P_{pS} \leq 0.32$ and $P_{n4} \leq 0.32$. This distribution includes 52,897 galaxies.}
\label{bulge_freen_dist}
\end{figure}

The choice of  a S\'ersic index value for the galaxy bulge profile has long been debated in the literature \citep{andredakis95,dejong96,balcells03,kormendy04,fisher10,laurikainen10}, and our very large sample of free $n_b$ + disk decompositions should at first glance offer some insight on this interesting question. However, the vast majority of the galaxies in our sample have images that do not have the required spatial resolution and/or signal-to-noise ratio (Section~\ref{ftest} and Figure~\ref{pps_pn4_btfn}). If we select galaxies for which both $P_{pS}$ and $P_{n4}$ are less or equal to 0.32, then we obtain a subsample of  almost 53,000 galaxies for which we have good enough images to study their bulge profile shape. Figure~\ref{bulge_freen_dist} shows the distribution of $n_b$ for this subsample. There are three important features to examine here. First, there is a peak at $n_b = 0.5$. We examined the distribution of galaxies with 0.5 $\leq n_b < 0.55$ in apparent bulge size $r_e$ and bulge ellipticity $e$, and this distribution showed that essentially all of these galaxies ($N \sim$ 6,700) were located in a peak at $e \sim$ 0.7 and $r_e \lesssim$ 1\arcsec-2\arcsec, i.e., 3-5 pixels. A combination of low $n_b$ and high $e$ values is expected when the fitting algorithm tries to make the bulge profile as flat and as elongated as possible to try to fit a bar or include off-center components (point sources, very close mergers, etc.). We visually inspected a subsample of galaxies in this peak using the SDSS SkyServer Object Explorer\footnote{http://cas.sdss.org/dr7/en/tools/explore/obj.asp} to confirm this expectation. Second, the $n_b$ distribution has a broad bump around $n_b = 5.5-6.0$. The de Vaucouleurs value of $n_b = 4$ is not preferred for this subsample. However, choosing $n_b=4$ for the entire sample is still a reasonable choice for the following reason. When the bulge S\'ersic index cannot be constrained due to spatial resolution and/or signal-to-noise limitations, its posterior probability distribution (which is fully mapped by GIM2D) will be uniformly flat between the minimum and maximum allowed values. Our allowed range of values was 0.5-8 based on previous studies of the S\'ersic index of spheroids. The median value of a flat posterior probability distribution (which we take to be the best-fit value) will therefore be around 4. Indeed, if we re-plot Figure~\ref{bulge_freen_dist} for the entire sample with no $F$-test selection, we see a very strong peak at $n_b=4$, but this peak reflects a lack of constraint on bulge profile rather than its actual shape. Third, there is an upturn in the $n_b$ distribution at $n_b \gtrsim$ 7.5. We again examined a plot of $r_e$ versus $e$ for these galaxies and found peaks at $e \sim 0$ and $e \sim 0.7$ with a uniform distribution in sizes over the range $r_e < 3$\arcsec. Visual inspection of galaxies at $e \sim 0$ and e $\sim$ 0.7 with smaller sizes ($r_e  \lesssim $ 0\arcsecpoint4) showed them to have nuclear, {\it on-center} sources. Galaxies at $e \sim 0$ and larger sizes ($r_e \sim $ 2$-$3\arcsec) did not exhibit any distinguishing characteristic as a group, and galaxies at $e \sim 0.7$ and larger sizes ($r_e \sim$ 2$-$3\arcsec) had a bar+point source configuration. Galaxies with low and high $n_b$ values in Figure~\ref{bulge_freen_dist} also have low and high $n_g$ values in Figure~\ref{dr7_fn_n4_pS_cmp_struct}. All of these different sub-classes of objects highlight the power of comparing different fitting models to identify different types of galaxy substructures.

\begin{figure*}
\begin{center}
\resizebox{\textwidth}{!}{\rotatebox{-90}{\includegraphics{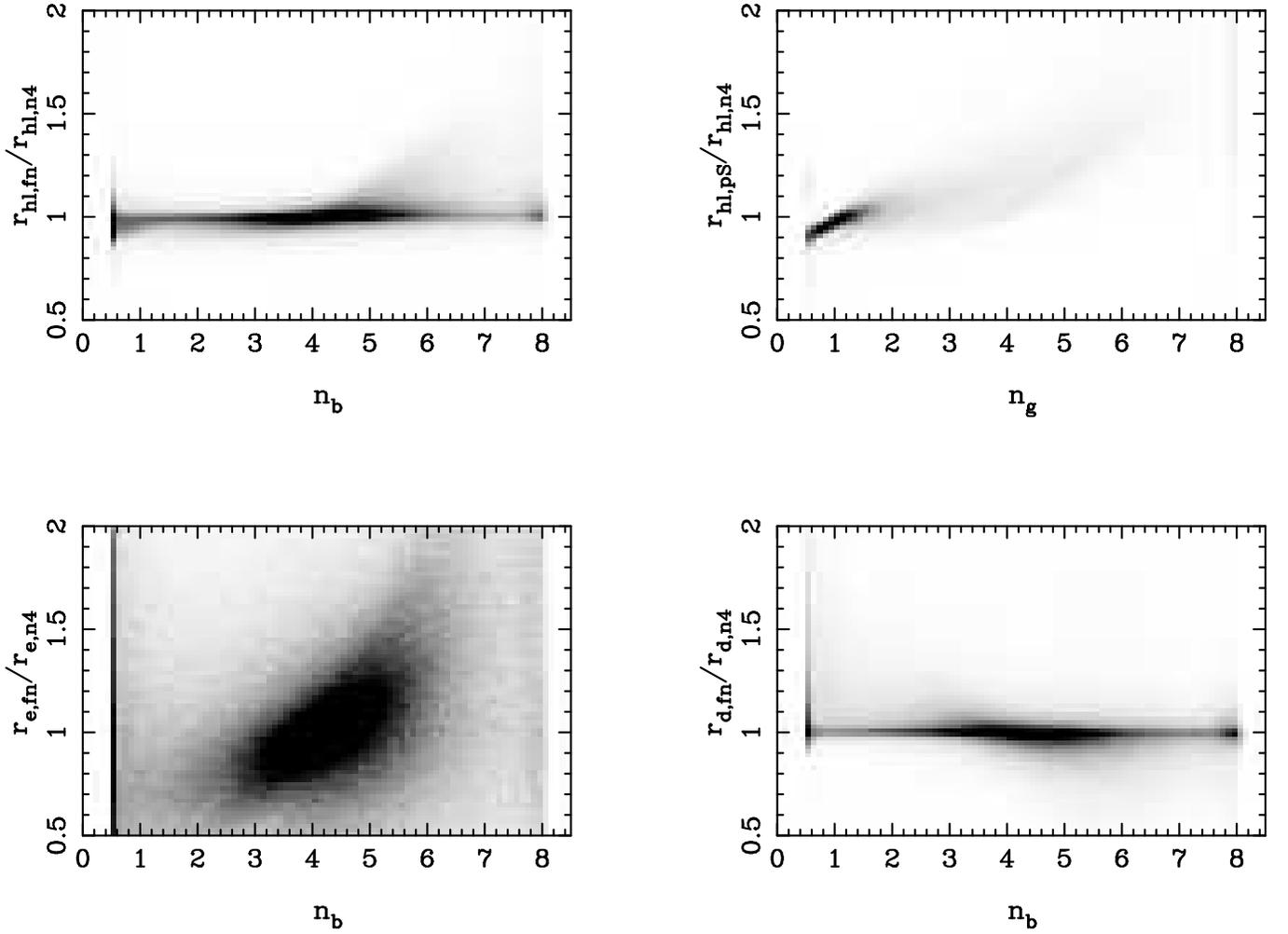}}}
\caption{Comparison of $r$-band GIM2D radii measured from three different fitting models.  {\it Top left:} Ratio of the galaxy half-light radii $r_{hl,fn}$ and $r_{hl,n4}$ from the free $n_b$ bulge + disk and  $n_b=4$ bulge + disk decompositions versus bulge S\'ersic index $n_b$. {\it Top right:} Ratio of the galaxy half-light radii $r_{hl,pS}$ and $r_{hl,n4}$ from the single component, pure S\'ersic fits and $n_b =4$ bulge + disk decompositions versus the galaxy S\'ersic index $n_g$. {\it Bottom left:} Ratio of the bulge effective radii $r_{e,fn}$ and $r_{e,n4}$ from the free $n_b$ bulge + disk and  $n_b =4$ bulge + disk decompositions versus bulge S\'ersic index $n_b$. {\it Bottom right:} Ratio of the disk scale lengths $r_{d,fn}$ and $r_{d,n4}$ from free $n_b$ bulge + disk and $n_b=4$ bulge + disk decompositions versus bulge S\'ersic index $n_b$. The greyscale represents the two-dimensional distribution of galaxies normalized by the total number of galaxies. The $x$-axis and $y$-axis bin sizes for the two-dimensional distributions were 0.1 and 0.02 respectively. The limits for the greyscale go from zero to 50$\%$ of the peak value of the normalized distribution.}
\label{dr7_fn_n4_pS_cmp_radii}
\end{center}
\end{figure*}

Galaxy half-light radii from $n_b = 4$ and free $n_b$ decompositions ($r_{hl,n4}$ and $r_{hl,fn}$ respectively) are quite consistent over the full range of allowed $n_b$ values (Figure~\ref{dr7_fn_n4_pS_cmp_radii}). Not surprisingly, galaxy half-light radii differ between pure Sersic ($r_{hl,pS}$) and bulge+disk fits at $n_g > 4$ with $r_{hl,pS}$ being 50$\%$ larger at $n_g = 8$. This is entirely due to the fact that the half-light radii are calculated by integrating best-fit models with different profiles in their outer wings. The bulge radii exhibit the expected shape from the well-known and strong covariance between $n_b$ and $r_e$, and the choice of $n_b$ will obviously have a significant impact on the measurements of $r_e$. On the other hand, it is very important to note that the disk scale length does {\it not} appear to be affected by the S\'ersic index of the bulge for the majority (80$\%$) of the galaxies in our sample, and that the scatter in the disk luminosity-size relation (Figure~\ref{disk-lumsize}) therefore does not depend on the choice of bulge S\'ersic index. This apparent lack of dependence of $r_d$ on $n_b$ may be due to the lack of constraint on bulge profile shape discussed earlier, but it may also be due to bulges being usually more compact than their disks. The covariance between measured bulge and disk parameters will be weaker in galaxies where the two components are more spatially distinct. More details on the disk luminosity-size distribution are given in Simard 2011, in preparation.

\subsubsection{S\'ersic Model Comparison with NYU Value-Added Catalog}\label{nyuvac}

The New York University Value-Added Galaxy Catalog \citep{blanton05a} provides S\'ersic model structural parameters for galaxies in the SDSS spectroscopic sample. The details and tests of the NYU S\'ersic measurements including artificial galaxy simulations are described in the appendix of \citet{blanton05b}. We matched objects in our pure S\'ersic structural catalog with objects in the NYU catalog using MJD, PLATEID and FIBERID for cross-identifications, and a match was found for 666,740 objects. 

\begin{figure*}
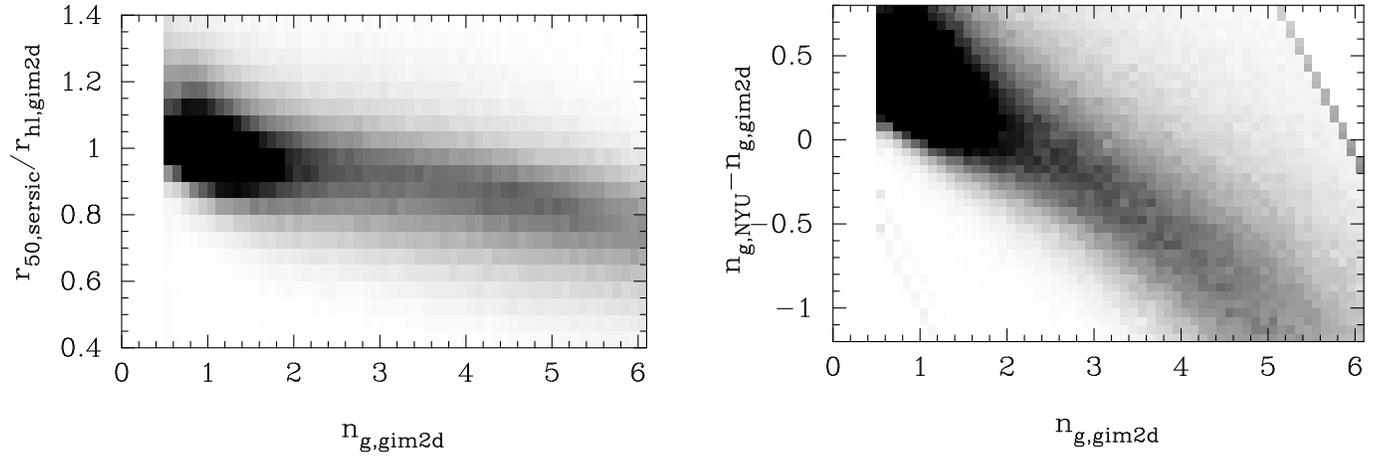

\includegraphics[angle=270,width=8.5cm]{fig17a.eps}
\hfill
\includegraphics[angle=270,width=8.5cm]{fig17b.eps}
\caption{Comparison between NYU and GIM2D pure S\'ersic structural parameters. {\it Left:} Ratio of the NYU and circularized GIM2D galaxy half-light radii $r_{50,sersic}$ and $r_{hl,gim2d}$ versus the GIM2D galaxy S\'ersic index $n_{g,gim2d}$.  {\it Right:} Difference between the NYU and GIM2D galaxy S\'ersic indices versus GIM2D galaxy S\'ersic index $n_{g,gim2d}$. The greyscale represents the two-dimensional distribution of galaxies normalized by the total number of galaxies. The $x$-axis and $y$-axis bin sizes for the two-dimensional distributions were 0.1 and 0.05 respectively. The limits for the greyscale go from zero to 30$\%$ of the peak value of the normalized distribution.}
\label{nyu-g2d_cmp}
\end{figure*}

Figure~\ref{nyu-g2d_cmp} shows the comparison between NYU and GIM2D galaxy S\'ersic half-light radii and indices.  The trend in half-light radius shown in the left-hand panel is fully consistent with the NYU simulations if $n_{g,gim2d}$ and $r_{hl,gim2d}$ are taken to be equivalent to the input (``true") values used for the NYU simulations ($n_{in}$ and $r_{50,in}$ respectively). The galaxy half-light radii from the NYU  fits are smaller by about 20$\%$ than both their input simulation values and the GIM2D values for objects with $n_g \geq 5$. There is an offset $\Delta n$ $\sim$ 0.3-0.4 between GIM2D and NYU S\'ersic indices at $n_g$ = 1. This offset depends on galaxy ellipticity: it increases from 0.2 at low $e$ ($<$ 0.1) to 0.5 at higher $e$  ($>$ 0.4). This dependence on ellipticity comes from the fact that the NYU profile fits were done on one-dimensional profiles extracted from two-dimensional images using {\it circular} annuli. The offset does not completely disappear even at low $e$ possibly as a result of the fact that the NYU fits were done on $r-$band images only whereas our fits were done simultaneously on $g-$ and $r-$band images, and a redder band will be more dominated by the redder, spheroidal (i.e., higher S\'ersic index) component of a galaxy. 

The comparison between the NYU and GIM2D S\'ersic parameters therefore shows good agreement given the differences in how the two sets of parameters were measured.

\subsection{Data Tables and Some Cautionary Notes}\label{bigtables}

The data quality metrics used in Section~\ref{QAmetrics} show that the GIM2D ``SIM+SEXTDEBL+GM2DBKG" dataset gives the most robust photometric results. The photometric data for this dataset are given in Tables~\ref{sdss_data_table_bd_n4}, \ref{sdss_data_table_bd_fn} and \ref{sdss_data_table_pS} for the $n_b=4$ bulge + disk, the free-$n_b$ bulge + disk and the pure S\'ersic decompositions respectively. Two sets of galaxy half-light radii are given in these tables. The semi-major half-light radius $R_{hl}$ of a galaxy was calculated by individually collapsing the bulge and disk components onto their respective major axes, adding these two one-dimensional profiles into a global galaxy profile and computing the half light radius of this summed one-dimensional profile. The circular half light radius $R_{chl}$ was computed by performing curve-of-growth photometry in circular apertures on the {\it intrinsic} (i.e., not PSF convolved) GIM2D best-fit model image of the galaxy. The relationship between the two kinds of half-light radius is easy to understand for a single component galaxy because it is given in this case by $R_{chl}$ = $R_{hl} \sqrt{1-e}$ where $e$ is the ellipticity of this component.

\begin{figure*}
\begin{center}
\includegraphics[angle=0,width=15cm]{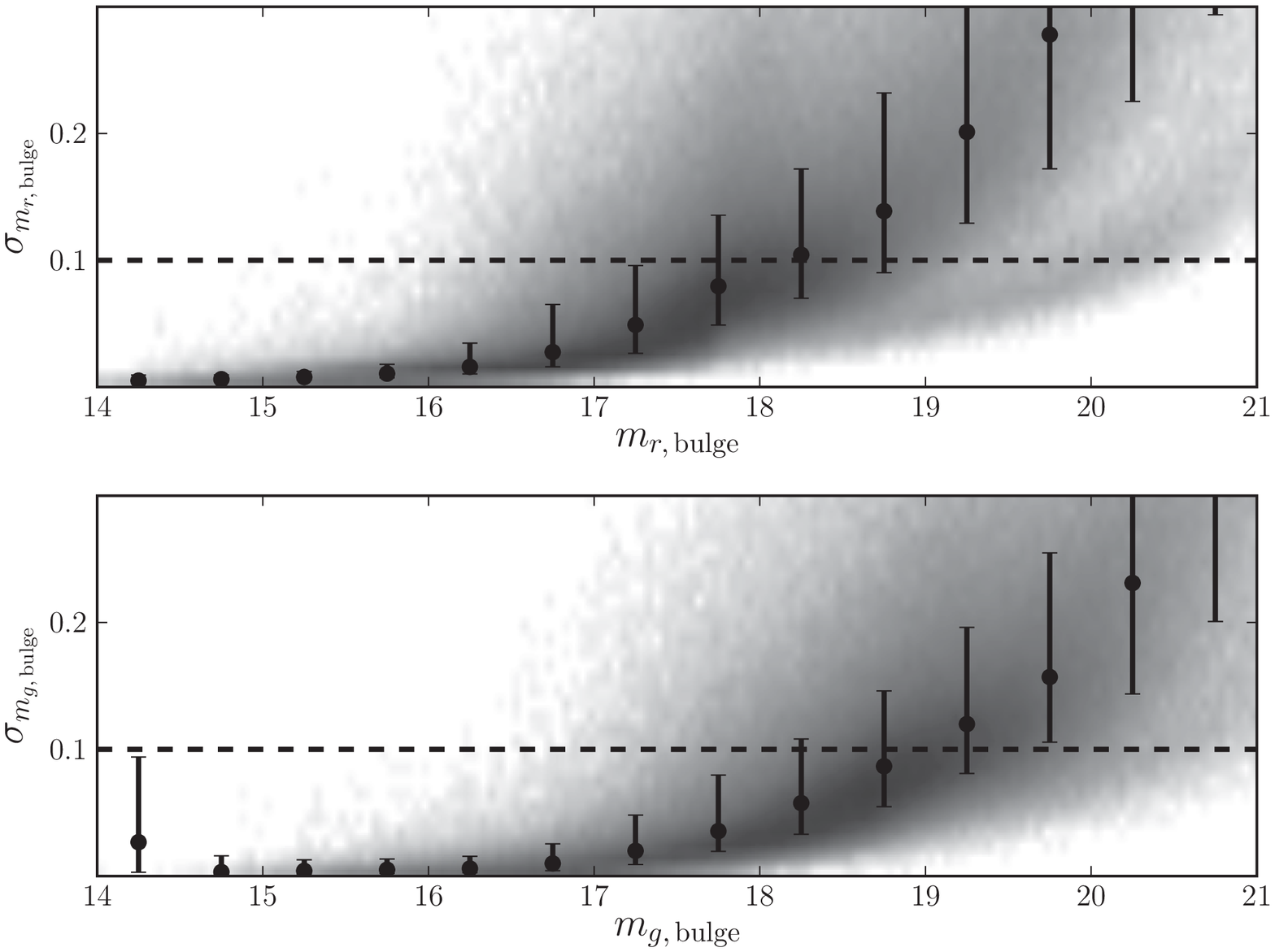}
\caption{GIM2D bulge $g$ and $r$ photometric errors as a function of bulge magnitude in n=4 bulge+disk decompositions. Bulge magnitudes were calculated using Equations~\ref{gmag-bulge} and~\ref{rmag-bulge}. Errors on total galaxy magnitudes and bulge fractions were propagated through these equations to obtain the photometric errors. The data points are median values, and the error bars are the 16th and 84th percentile values.}
\label{err_mag_bulge}
\end{center}
\end{figure*}

\begin{figure*}
\begin{center}
\includegraphics[angle=0,width=15cm]{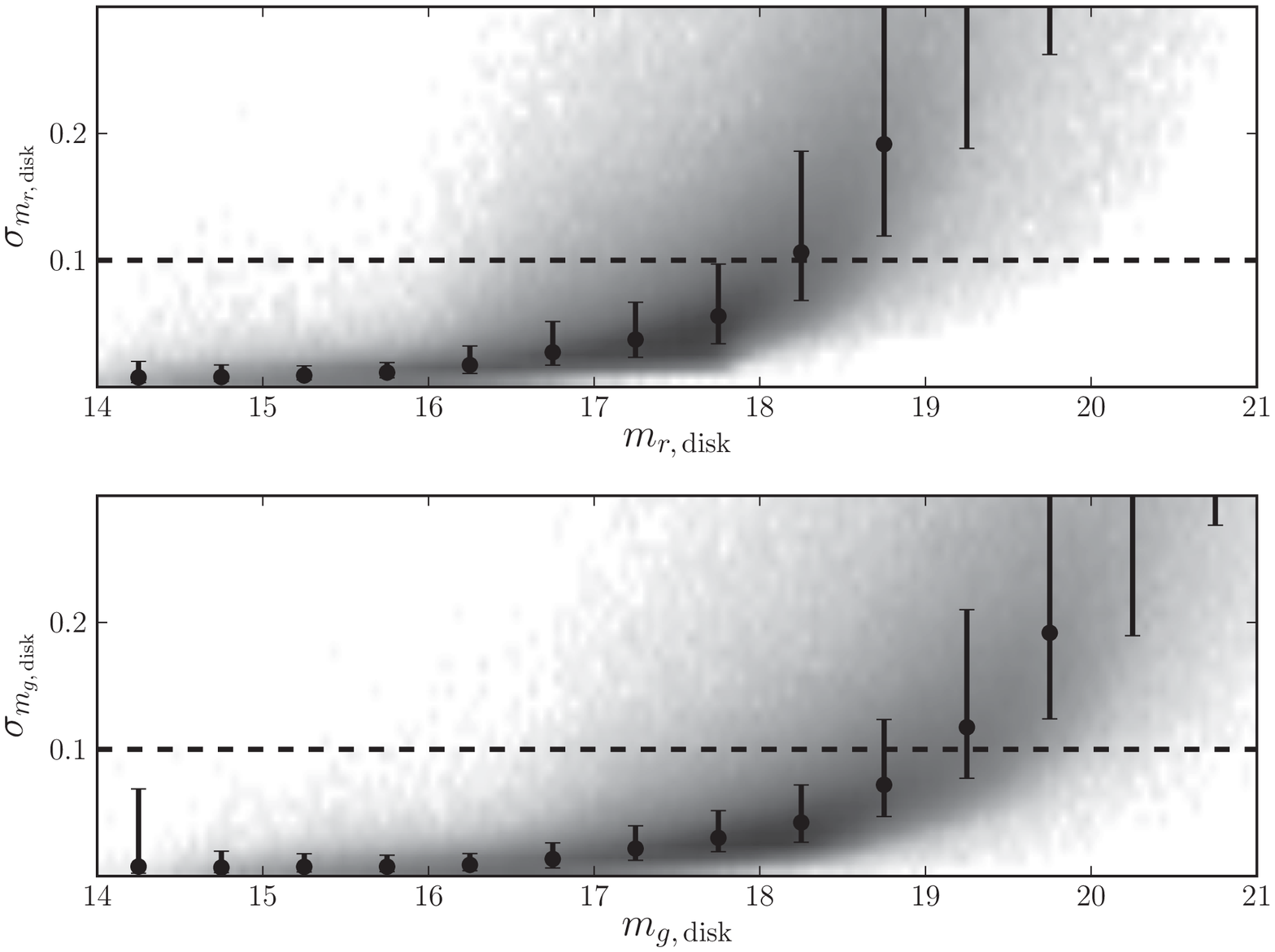}
\caption{GIM2D disk $g$ and $r$ photometric errors as a function of disk magnitude in n=4 bulge + disk. Disk magnitudes were calculated using Equations~\ref{gmag-disk} and~\ref{rmag-disk}. Errors on total galaxy magnitudes and bulge fractions were propagated through these equations to obtain the photometric errors. The data points are median values, and the error bars are the 16th and 84th percentile values.}
\label{err_mag_disk}
\end{center}
\end{figure*}

The best way to use the data listed in the structural parameter tables is to consider bulges and disks as separate galaxy sub-populations overlapping on the sky. As one would expect, bulge structural parameters are more reliable for brighter bulges (Figure~\ref{err_mag_bulge}), and disk structural parameters are more reliable for brighter disks (Figure~\ref{err_mag_disk}). The reliability of the structural parameters of a given subcomponent is largely independent of the other subcomponent (e.g.,  Figure~\ref{dr7_fn_n4_pS_cmp_radii}). One should not study bulge or disk properties on the basis of a simple selection cut on bulge fraction. The bulge of a bright, low $B/T$ galaxy can still be brighter than the bulge of a faint, high $B/T$ galaxy. Bulge and disk subsamples should be selected on the basis of their magnitudes. 

Another important note of caution is related to the use of bulge fraction cuts to select early-type galaxy subsamples. As many previous studies have shown \citep{im02,mcintosh02,tran03,blakeslee06,simard09}, early-type galaxies should be selected using both bulge fraction {\it and} image smoothness. For example, a nuclear starburst in a relatively irregular galaxy would yield a high bulge fraction, and such a galaxy would be erroneously classified as an early-type if bulge fraction were the sole selection criterion. The agreement between visual and quantitative classification of galaxies has been shown to be excellent when both parameters are used \citep{simard09}.

Finally, internal dust should also be considered as a potential source of bias when selecting samples of bulges and/or disks for study. Disks are known to be dusty \citep[e.g.,][and references therein]{driver07}, and this dust can broaden the intrinsic disk/bulge luminosity-size or luminosity-color distributions. One can deal with this bias by either selecting galaxies with face-on disks or applying empirical internal dust extinction corrections. These corrections can be difficult to characterize, but the large number of galaxies included in the catalogs presented here offers the opportunity of deriving these corrections as needed by various science programs (Simard 2011, in preparation).

\section{Summary}\label{summary}
We have performed bulge+disk decompositions for 1.12 million galaxies in the Legacy area of the SDSS Data Release 7. Four decomposition procedures were used, and one of these procedures clearly produced more robust structural parameters, magnitude and colors when looking at three science-based data quality assurance metrics. The most reliable procedure included the following three important steps:
\begin{itemize}
\item GIM2D-based sky background level determination
\item SExtractor object deblending
\item Simultaneous bulge+disk decomposition in $g$ and $r$
\end{itemize}
We also used three different fitting models: a $n_b=4$ bulge + disk model, a free-$n_b$ bulge + disk model and a pure S\'ersic model, and we provide a detailed comparison between the measured structural parameters from these fitting models with a mean to select the appropriate model for a given galaxy using the $F$-statistic. This comparison highlights the importance of this selection for a given science goal. Pure S\'ersic model fits might be better suited to studies of global galaxy colours whereas a cut on the $F$-test probability $P_{pS}$ to select galaxies for which a bulge+disk model was required would be better for studies of bulge/disk colours. Using a low cut on both $P_{pS}$ and $P_{pS}$ probabilities would select relatively small but very robust samples of bulge+disk decompositions. Full catalogs of structural parameters are included here with important cautionary notes on how to select bulge and/or disk subsamples, how to select different morphological types, the issue of bars and the issue of internal dust. These catalogs should provide an extensive comparison set for a wide range of observational and theoretical studies of galaxies.

\acknowledgments

LS, SLE and DRP gratefully acknowledge financial support from Discovery Grants through the Natural Science and Engineering Research
Council of Canada. This research made use of a University of Victoria computing facility funded by grants from the Canadian Foundation for
Innovation and the British Columbia Knowledge and Development Fund. We thank the system administrators of this facility for their flawless support.
 Funding for the creation and distribution of the
SDSS Archive has been provided by the Alfred P. Sloan Foundation, the
Participating Institutions, the National Aeronautics and Space
Administration, the National Science Foundation, the U.S. Department
of Energy, the Japanese Monbukagakusho, and the Max Planck
Society. The SDSS Web site is http://www.sdss.org/.The SDSS is managed
by the Astrophysical Research Consortium (ARC) for the Participating
Institutions. The Participating Institutions are The University of
Chicago, Fermilab, the Institute for Advanced Study, the Japan
Participation Group, The Johns Hopkins University, the Korean
Scientist Group, Los Alamos National Laboratory, the
Max-Planck-Institute for Astronomy (MPIA), the Max-Planck-Institute
for Astrophysics (MPA), New Mexico State University, University of
Pittsburgh, University of Portsmouth, Princeton University, the United
States Naval Observatory, and the University of Washington.

\clearpage
\begin{deluxetable}{ll}
\tablecaption{SDSS structural parameters from $n_b$=4 bulge + disk decompositions (Table available in online electronic version)\label{sdss_data_table_bd_n4}}
\tabletypesize{\tiny}
\tablewidth{0pt}
\tablehead{\colhead{Column}& \colhead{Description}\\
\colhead{Name}& }
\startdata
ObjID & SDSS Object ID \\ 
$z$ & SDSS Redshift (Spectroscopic if available. Photometric otherwise)\\ 
SpecClass & SDSS SpecClass value (set to $-$1 if $z$ is photometric or $-$2 if no redshift available at all)\\
Scale & Physical scale in arcsec/kpc at redshift $z$\\
$V_{max}$ & Galaxy volume correction in Mpc$^3$ (Equation~\ref{vmax-integral})\\ 
$g_{g2d}$ & $g$-band apparent magnitude of GIM2D output B+D model (Equation~\ref{g2dmag})\\ 
$r_{g2d}$ & $r$-band apparent magnitude of GIM2D output B+D model (Equation~\ref{g2dmag})\\
$g_{g2d,f}$ & $g$-band apparent fiber magnitude of output B+D model \\ 
$r_{g2d,f}$ & $r$-band apparent fiber magnitude of output B+D model\\
$\Delta$(fiber color) & Delta fiber color defined as $(g-r)_{gim2d,fiber} - (g-r)_{SDSS,fiber}$ (set to $-$99.99 if no SDSS fiber magnitudes available)\\
$(B/T)_g$ & $g$-band bulge fraction\\ 
$(B/T)_r$ & $r$-band bulge fraction\\ 
$(B/T)_{g,f}$ & $g$-band fiber bulge fraction\\
$(B/T)_{r,f}$ & $r$-band fiber bulge fraction\\
$R_{hl,g}$ & $g$-band galaxy semi-major axis, half-light radius in kiloparsecs\\
$R_{hl,r}$ & $r$-band galaxy semi-major axis, half-light radius in kiloparsecs\\
$R_{chl,g}$ & $g$-band galaxy circular half-light radius in kiloparsecs\\
$R_{chl,r}$ & $r$-band galaxy circular half-light radius in kiloparsecs\\
$R_e$ & Bulge semi-major effective radius in kiloparcsecs (Equation~\ref{ang-diam})\\ 
$e$ & Bulge ellipticity ($e \equiv  1 - b/a$, e = 0 for a circular bulge)\\ 
$\phi_b$ & Bulge position angle in degrees (measured clockwise from the $+y$ axis of SDSS images)\\ 
$R_d$ & Exponential disk scale length in kiloparsecs (Equation~\ref{ang-diam})\\ 
$i$ & Disk inclination angle in degrees ($i \equiv$ 0 for a face-on disk) \\ 
$\phi_d$ & Disk position angle in degrees (measured clockwise from the $+y$ axis of SDSS images)\\
$(dx)_g$ & B+D model center offset from column position given by {\tt colc\_g} on SDSS corrected $g$-band image (arcsec)\\
$(dy)_g$ & B+D model center offset from row position given by {\tt rowc\_g} on SDSS corrected $g$-band image (arcsec)\\
$(dx)_r$ & B+D model center offset from column position given by {\tt colc\_r} on SDSS corrected $r$-band image (arcsec)\\
$(dy)_r$ & B+D model center offset from row position given by {\tt rowc\_r} on SDSS corrected $r$-band image (arcsec)\\
$S2_g$ & $g$-band image smoothness parameter (as defined in \citet{simard09})\\ 
$S2_r$ & $r$-band image smoothness parameter  (as defined in \citet{simard09})\\
$M_{g,g}$ & $g$-band GIM2D galaxy rest-frame,  absolute magnitude (Equation~\ref{gmag-galaxy})\\
$M_{g,b}$ & $g$-band GIM2D bulge rest-frame,  absolute magnitude (Equation~\ref{gmag-bulge})\\ 
$M_{g,d}$ & $g$-band GIM2D disk rest-frame,  absolute magnitude (Equation~\ref{gmag-disk})\\ 
$M_{r,g}$ & $r$-band GIM2D galaxy rest-frame,  absolute magnitude (Equation~\ref{rmag-galaxy})\\
$M_{r,b}$ & $r$-band GIM2D bulge rest-frame,  absolute magnitude (Equation~\ref{rmag-bulge})\\ 
$M_{r,d}$ & $r$-band GIM2D disk rest-frame,  absolute magnitude (Equation~\ref{rmag-disk})\\
$n_b$ & Bulge S\'ersic index\\ 
$P_{pS}$ & $F$-test probability that a B+D model is {\it not} required compared to a pure S\'ersic model\\
\enddata
\end{deluxetable}

\clearpage
\begin{deluxetable}{ll}
\tablecaption{SDSS structural parameters from free $n_b$ bulge + disk decompositions (Table available in online electronic version)\label{sdss_data_table_bd_fn}}
\tabletypesize{\tiny}
\tablewidth{0pt}
\tablehead{\colhead{Column}& \colhead{Description}\\
\colhead{Name}& }
\startdata
ObjID & SDSS Object ID \\ 
$z$ & SDSS Redshift (Spectroscopic if available. Photometric otherwise)\\ 
SpecClass & SDSS SpecClass value (set to $-$1 if $z$ is photometric or $-$2 if no redshift available at all)\\
Scale & Physical scale in arcsec/kpc at redshift $z$\\
$V_{max}$ & Galaxy volume correction in Mpc$^3$ (Equation~\ref{vmax-integral})\\ 
$g_{g2d}$ & $g$-band apparent magnitude of GIM2D output B+D model (Equation~\ref{g2dmag})\\ 
$r_{g2d}$ & $r$-band apparent magnitude of GIM2D output B+D model (Equation~\ref{g2dmag})\\
$g_{g2d,f}$ & $g$-band apparent fiber magnitude of output B+D model \\ 
$r_{g2d,f}$ & $r$-band apparent fiber magnitude of output B+D model\\
$\Delta$(fiber color) & Delta fiber color defined as $(g-r)_{gim2d,fiber} - (g-r)_{SDSS,fiber}$ (set to $-$99.99 if no SDSS fiber magnitudes available)\\
$(B/T)_g$ & $g$-band bulge fraction\\ 
$(B/T)_r$ & $r$-band bulge fraction\\ 
$(B/T)_{g,f}$ & $g$-band fiber bulge fraction\\
$(B/T)_{r,f}$ & $r$-band fiber bulge fraction\\
$R_{hl,g}$ & $g$-band galaxy semi-major axis, half-light radius in kiloparsecs\\
$R_{hl,r}$ & $r$-band galaxy semi-major axis, half-light radius in kiloparsecs\\
$R_{chl,g}$ & $g$-band galaxy circular half-light radius in kiloparsecs\\
$R_{chl,r}$ & $r$-band galaxy circular half-light radius in kiloparsecs\\
$R_e$ & Bulge semi-major effective radius in kiloparcsecs (Equation~\ref{ang-diam})\\ 
$e$ & Bulge ellipticity ($e \equiv  1 - b/a$, e = 0 for a circular bulge)\\ 
$\phi_b$ & Bulge position angle in degrees (measured clockwise from the $+y$ axis of SDSS images)\\ 
$R_d$ & Exponential disk scale length in kiloparsecs (Equation~\ref{ang-diam})\\ 
$i$ & Disk inclination angle in degrees ($i \equiv$ 0 for a face-on disk) \\ 
$\phi_d$ & Disk position angle in degrees (measured clockwise from the $+y$ axis of SDSS images)\\
$(dx)_g$ & B+D model center offset from column position given by {\tt colc\_g} on SDSS corrected $g$-band image (arcsec)\\
$(dy)_g$ & B+D model center offset from row position given by {\tt rowc\_g} on SDSS corrected $g$-band image (arcsec)\\
$(dx)_r$ & B+D model center offset from column position given by {\tt colc\_r} on SDSS corrected $r$-band image (arcsec)\\
$(dy)_r$ & B+D model center offset from row position given by {\tt rowc\_r} on SDSS corrected $r$-band image (arcsec)\\
$S2_g$ & $g$-band image smoothness parameter (as defined in \citet{simard09})\\ 
$S2_r$ & $r$-band image smoothness parameter  (as defined in \citet{simard09})\\
$M_{g,g}$ & $g$-band GIM2D galaxy rest-frame,  absolute magnitude (Equation~\ref{gmag-galaxy})\\
$M_{g,b}$ & $g$-band GIM2D bulge rest-frame,  absolute magnitude (Equation~\ref{gmag-bulge})\\ 
$M_{g,d}$ & $g$-band GIM2D disk rest-frame,  absolute magnitude (Equation~\ref{gmag-disk})\\ 
$M_{r,g}$ & $r$-band GIM2D galaxy rest-frame,  absolute magnitude (Equation~\ref{rmag-galaxy})\\
$M_{r,b}$ & $r$-band GIM2D bulge rest-frame,  absolute magnitude (Equation~\ref{rmag-bulge})\\ 
$M_{r,d}$ & $r$-band GIM2D disk rest-frame,  absolute magnitude (Equation~\ref{rmag-disk})\\
$n_b$ & Bulge S\'ersic index\\ 
$P_{pS}$ & $F$-test probability that a B+D model is {\it not} required compared to a pure S\'ersic model\\
$P_{n4}$ & $F$-test probability that a free $n_b$ B+D model is {\it not} required compared to a fixed $n_b$=4 B+D model\\
\enddata
\end{deluxetable}

\clearpage
\begin{deluxetable}{ll}
\tablecaption{SDSS structural parameters from pure S\'ersic decompositions (Table available in online electronic version)\label{sdss_data_table_pS}}
\tabletypesize{\tiny}
\tablewidth{0pt}
\tablehead{\colhead{Column}& \colhead{Description}\\
\colhead{Name}& }
\startdata
ObjID & SDSS Object ID \\ 
$z$ & SDSS Redshift (Spectroscopic if available. Photometric otherwise)\\ 
SpecClass & SDSS SpecClass value (set to $-$1 if $z$ is photometric or $-$2 if no redshift available at all))\\
Scale & Physical scale in arcsec/kpc at redshift $z$\\
$V_{max}$ & Galaxy volume correction in Mpc$^3$ (Equation~\ref{vmax-integral})\\ 
$g_{g2d}$ & $g$-band apparent magnitude of GIM2D output pure S\'ersic model (Equation~\ref{g2dmag})\\ 
$r_{g2d}$ & $r$-band apparent magnitude of GIM2D output pure S\'ersic model (Equation~\ref{g2dmag})\\
$g_{g2d,f}$ & $g$-band apparent fiber magnitude of output pure S\'ersic model \\ 
$r_{g2d,f}$ & $r$-band apparent fiber magnitude of output pure S\'ersic model\\
$\Delta$(fiber color) & Delta fiber color defined as $(g-r)_{gim2d,fiber} - (g-r)_{SDSS,fiber}$ (set to $-$99.99 if no SDSS fiber magnitudes available)\\
$R_{hl,g}$ & $g$-band galaxy semi-major axis, half-light radius in kiloparsecs\\
$R_{hl,r}$ & $r$-band galaxy semi-major axis, half-light radius in kiloparsecs\\
$R_{chl,g}$ & $g$-band galaxy circular half-light radius in kiloparsecs\\
$R_{chl,r}$ & $r$-band galaxy circular half-light radius in kiloparsecs\\
$e$ & Galaxy ellipticity ($e \equiv  1 - b/a$, e = 0 for a circular galaxy)\\ 
$\phi_b$ & Galaxy position angle in degrees (measured clockwise from the $+y$ axis of SDSS images)\\ 
$(dx)_g$ & Pure S\'ersic model center offset from column position given by {\tt colc\_g} on SDSS corrected $g$-band image (arcsec)\\
$(dy)_g$ & Pure S\'ersic model center offset from row position given by {\tt rowc\_g} on SDSS corrected $g$-band image (arcsec)\\
$(dx)_r$ & Pure S\'ersic  model center offset from column position given by {\tt colc\_r} on SDSS corrected $r$-band image (arcsec)\\
$(dy)_r$ & Pure S\'ersic model center offset from row position given by {\tt rowc\_r} on SDSS corrected $r$-band image (arcsec)\\
$S2_g$ & $g$-band image smoothness parameter (as defined in \citet{simard09})\\ 
$S2_r$ & $r$-band image smoothness parameter  (as defined in \citet{simard09})\\
$M_{g,g}$ & $g$-band GIM2D galaxy rest-frame,  absolute magnitude (Equation~\ref{gmag-galaxy})\\ 
$M_{r,g}$ & $r$-band GIM2D galaxy rest-frame,  absolute magnitude (Equation~\ref{rmag-galaxy})\\ 
$n_g$ & Galaxy S\'ersic index \\
\enddata
\end{deluxetable}

\end{document}